\documentclass[
reprint,
 amsmath,amssymb,
 aps,
floatfix,
]{revtex4-2}

\usepackage{graphicx}
\usepackage{dcolumn}
\usepackage{bm}
\usepackage{comment}
\usepackage{url}
\begin{document}


\title{A Systematic Study of Projection Biases in Weak Lensing Analysis}

\author{P.R.V.~Chintalapati}
\email{raj.ch90@gmail.com}
 \affiliation{Northern Illinois University}
 \affiliation{Fermi National Accelerator Laboratory, P.O. Box 500, Batavia, IL 60510, USA.}
\author{G.~Gutierrez}
 \email{gaston@fnal.gov}
\affiliation{Fermi National Accelerator Laboratory, P.O. Box 500, Batavia, IL 60510, USA.}
\author{M.H.L.S.~Wang}
 \email{mwang@fnal.gov}
\affiliation{Fermi National Accelerator Laboratory, P.O. Box 500, Batavia, IL 60510, USA.}

\date{\today}

\begin{abstract}
We present a systematic study of projection biases in the weak lensing analysis of the first year of data from the Dark Energy Survey (DES) experiment.  In the analysis we used a $\Lambda$CDM model and three two-point correlation functions.  We show that these biases are a consequence of projecting, or marginalizing, over parameters like $h$, $\Omega_b$, $n_s$ and $\Omega_\nu h^2$ that are both poorly constrained and correlated with the parameters of interest like $\Omega_m$, $\sigma_8$ and $S_8$.   Covering the relevant parameter space we show that the projection biases are a function of where the true values of the poorly constrained parameters lie with respect to the parameter priors.  For example, biases in the position of the posteriors can exceed the 1.5$\sigma$ level if the true values of $h$ and $n_s$ are close to the top of the prior's range and the true values of $\Omega_b$ and $\Omega_\nu h^2$ are close to the bottom of the range of their priors.  We also show that in some cases the 1D credible intervals can be over-specified by as much as 30\% and coverage can be as low as 27\%.  Finally we estimate these projection biases for the analysis of three and six years worth of DES data.
\end{abstract}

\maketitle


\section{Introduction}\label{sec:introduction}

Since its first statistically significant observations two decades ago \cite{Sloan_WL,Nature_2000_WL,MNRAS_2000_WL,A&A_2000_WL}, weak lensing has evolved to become a powerful tool in determining the values of important cosmological parameters.  The recent results of the Dark Energy Survey (DES) experiment's first three years of data \cite{DES_Y3_1x2pt-a,DES_Y3_1x2pt-b,DES_Y3_2x2pt-a,DES_Y3_2x2pt-b,DES_Y3_3x2pt}, the Hyper Suprime-Cam Subaru Strategic Program's (HSC) first year of data \cite{HSC_2020} and the Kilo Degree Survey's fourth data release (KiDS) \cite{KiDS-1000_2021_1x2pt,KiDS-1000_2021_3x2pt,KiDS-1000_2021_beyondLCDM}, clearly show the power of weak lensing in measuring the normalization of the power spectrum $\sigma_8$, the total relative matter density $\Omega_m$, and their combination $S_8 = \sigma_8 \sqrt{\Omega_m / 0.3}$.  The already strong constraining power of weak lensing will rapidly increase with the analysis of the recently released shear catalog with three years of data from HSC \cite{HSC_2021_3YearCatalog}, and the upcoming analysis of the full six years of data from DES.  This constraining power will increase further with the upcoming ``Stage IV" surveys like the Euclid Space Telescope \footnote{Euclid: https://www.euclid-ec.org} expected to launch in 2022, the Vera Rubin Observatory (LSST) \footnote{LSST: https://www.lsst.org} expected to start operating in mid-2020s, and the Nancy Roman Space Telescope \footnote{NRST: https://roman.gsfc.nasa.gov} expected to launch also in the mid 2020s.

All these weak lensing analysis use Bayesian methods to extract information about cosmological or physical parameters.  They use the peaks, or averages, of the one or two dimensional posteriors to quote most likely values and 68\% or 95\% areas under the posteriors to quote credible intervals (or regions in the multidimensional case).  In the KiDS publications of their fourth data release, they correctly name the Bayesian intervals extracted from the posteriors as ``credible intervals" \cite{KiDS-1000_2021_1x2pt,KiDS-1000_2021_3x2pt,KiDS-1000_2021_beyondLCDM}.  On the other hand both DES \cite{DES_Y3_1x2pt-a,DES_Y3_1x2pt-b,DES_Y3_2x2pt-a,DES_Y3_2x2pt-b,DES_Y3_3x2pt} and HSC \cite{HSC_2020} loosely refer to their intervals as ``confidence intervals" (or C.L. when referring to the limits of the intervals).  A confidence interval is a frequentist concept that is associated with the idea of coverage probability.  That is, in a 68.27\% confidence interval there is a 68.27\% probability that the true value will be inside that interval, and if this is not the case it is said that the confidence interval doesn't satisfy the coverage condition.  Credible intervals on the other hand are interpreted as assigning a certain ``degree of belief" (e.g. 68.27\%) that the true value will be inside the credible interval.  More often than not Bayesian statistics is the only practical way of calculating confidence intervals, and in these cases we think it is important to check the ``coverage" of these intervals because the community at large usually interprets errors in a frequentist way.  For example, 1.5$\sigma$ discrepancies are ignored because they have a 13.4\% probability of happening, while a 3$\sigma$ difference between two measurements is interpreted as evidence that the discrepancy could be real because there is only a 0.26\% probability for this to happen.  In this paper we will address the question of coverage of the credible intervals calculated in weak lensing analysis for a wide range of cosmological parameters.

The importance of understanding any form of systematic and/or statistical biases present in weak lensing analyses cannot be overstated.  And the importance of this understanding increases as the statistics and therefore the constraining power of the weak lensing experiments continues to improve.  Different kinds of simulations are used to test the performance of these analyses.  For example, full N-body simulations have been used to assess biases in the shear measurement of galaxies or the errors in the photometric measurements, as well as checking if the approximations made in the analysis theory, like the calculation of the non-linear power spectrum or the treatment of intrinsic alignment, introduces systematic errors in the measurements (see for example reference \cite{DES_Y3_3x2pt_MC} for DES and \cite{KiDS_modify} for KiDS).  Faster simulations have been used to check, for example, the stability of the results to different aspects of the theory, or to validate covariance matrix models, or to study the effects of patchy sky coverage (see for example reference \cite{KiDS-1000_2021_MC} for KiDS and \cite{Krause_2017} for DES).

To check the coverage probability of a confidence interval, or region, we imagine performing the same experiment an infinite, or very large, number of times and then counting how often the true values of the parameters falls inside the specified confidence intervals.  Ideally we would do this generating a very large number of full N-body simulations and then running them through the analysis pipeline.  Using full body simulations would tests the full set of biases, e.g. biases due to measurement errors, theory approximations, shear produced by intrinsic alignment, or biases due to the way the credible intervals are calculated.  Unfortunately a program like this requires inordinate amounts a CPU time, so in a first step we will restrict ourselves to assuming that there are no measurement errors and that the analysis theory perfectly reflects nature.  Furthermore we will assume that the covariance matrix used to fit the two point correlation functions in weak lensing analysis correctly represents the fluctuations arising from performing the same experiment an infinite number of times.
Therefore in this paper we will study the question of coverage of weak lensing credible intervals by, 1) calculating a ``synthetic data vector" using the analysis theory to generate the two-point correlation functions used in the analysis, 2) fluctuating this synthetic data vector using the data covariance matrix, 3) using this fluctuated synthetic data vector as input to the weak lensing analysis, and 4) repeating steps 2 and 3 a large number of times while checking at every iteration the measurements with the theory values.  This clearly eliminates all questions about the validity of the theory and the existence of measurement errors.  We will refer to every iteration as an ``experiment" or ``pseudo-experiment" and to the entire process as ``doing ensemble tests".

Ensemble tests have the advantage of uncovering biases both in the shift of the posterior distributions and in the width of credible intervals, but they are still very CPU intensive due to the hundreds of jobs that need to be run in order to determine biases with enough statistical significance \footnote{The studies presented in this paper required about six million hours of CPU time, 83\% of which was used in the ensemble tests}.  We will show that just running the synthetic data vector (no fluctuations, just ``one experiment") gives biases in the shifts of the posteriors that are consistent with the ensemble tests.  We will also show that making some reasonable extrapolations from the width of the pull distributions obtained in the ensemble tests it is possible to get very good estimates of the credible interval's coverage for runs in which only the synthetic data vector was analyzed.  As we will see this is possible because, for example, most of the under-coverage of the credible intervals is due to biases in the position of the posteriors distributions.
Therefore, in this publication, we performed studies with ensemble tests for only five different combinations of cosmological parameters, or five fiducial cosmologies, and then used synthetic data vectors to explore the rest of the parameter space.  Synthetic data vectors have been used, for example, to study biases in the recently released DES weak lensing analysis \cite{DES_Y3_3x2pt_validation}.  In this reference only one fiducial cosmology was used, but the authors correctly point out that these projection effects depend ``on the parameterization and prior choices, the underlying parameter values, as well as the data's constraining power."   
In this paper we cover all the relevant parameter space and simulate an increase in the data's constraining power, and even though we leave the priors fixed, studying the biases for different position of the parameters inside their priors gives a very good idea of what would happens if one were to change these priors.  
In summary we will quantitatively show that, 1) the "projection biases" are a consequence of projecting, or marginalizing (read ``integrating" in both cases), the posterior distributions over parameters that are correlated with the parameters of interest and whose distributions are wider than the range of their priors, and 2) show that these biases are a function of where the true value of these poorly constrained parameters lie with respect to their prior intervals.

To perform the systematic study of projection biases presented in this paper, we used the publicly available code and data of the DES weak lensing analysis of their first year of data \cite{DES_Y1_PRD}.  We restricted ourselves to using a $\Lambda$CDM cosmology and the so called 3x2pt analysis, which uses all three two-point correlation functions, termed galaxy clustering, galaxy-galaxy lensing, and cosmic shear.  The bias studies of cosmic shear only (1x2pt), the combination of galaxy clustering and galaxy-galaxy lensing (2x2pt) and the effect of using a $w$CDM cosmology will be presented in a future paper.

In Section \ref{sec:Y1_analysis}, we summarize the model used in the analysis and give details of the public code that we used in our studies.  In Section \ref{sec:ProjectionBias}, we describe a simple but very instructive example about how the projection biases arise when projecting over parameters that are not well constrained, and show our main results on the biases in the position of the posterior distributions and the width and coverage of the credible intervals.  In Section \ref{sec:LargerStatistics}, we make a simple extrapolation to predict the projection bias behaviour with larger statistics, by dividing the data covariance matrix by factors corresponding to a DES analysis of three and six years worth of data. 
In this last section we will show that if priors are not tightened, many of the projection biases disappear with large increases in statistics, or reductions in the covariance matrix.  Experimenters always face the compromise of using wide prior ranges to make independent measurements or using external data to tighten the range of poorly constrained parameters in order to reduce their error bars.  Since the field advances in unison, experimenters will always operate somewhere in the middle and, in our opinion, will always be forced to understand the type of biases described in this paper.

Weak lensing provides tight constraints mostly on $\Omega_m$, $\sigma_8$, and $S_8$, so most of the plots and projection biases shown in this paper will concentrate on those three parameters.


\section{The DES Y1 \lowercase{3x2pt} WL Analysis}\label{sec:Y1_analysis}

For the projection bias studies presented in this paper, we will use the code that was developed by the DES collaboration to analyze the weak lensing (WL) data collected during the first year of observing (Y1) \cite{DES_Y1_PRD}.  That code and the data are publicly available in CosmoSIS \cite{Cosmosis_paper,Cosmosis_program}.  Furthermore, we will concentrate on studying the projection biases in the case in which $\Lambda$CDM is the fiducial cosmology, and the cosmological parameters are extracted by fitting to all three two-point correlation functions: 1) Galaxy clustering, which correlates the galaxy number densities along two different lines of sight (LOS), 2) Galaxy-galaxy lensing, that correlates the galaxy number density along a LOS with the galaxy shear along another LOS, and 3) Cosmic shear, which correlates galaxy shears along two different LOS.  In all cases, the two-point correlations are given as a function of the angle $\theta$ separating the two LOS, and in bins of redshift along the LOS.  In Subsection \ref{subsec:Y1_model}, we will give a short description of how the two-point correlations are calculated. In Subsection \ref{subsec:Y1_code}, we will describe the code that was used and how we selected the ``true" or nominal values of the parameters for our studies.

\subsection{The Model}\label{subsec:Y1_model}

In the weak lensing analysis of the data collected during the first year of running, DES used the flat sky and Limber approximations to calculate two-point correlation functions.  In this section, we will give a brief account of how the two-point correlation functions were calculated with the main purpose of making clear how the parameters listed in Table \ref{tab:parameters} enter in the analysis.

\begin{table}[hbt]
\caption{\label{tab:parameters}%
Table showing the parameters that were used in the DES Year 1 3x2pt WL analysis \cite{DES_Y1_PRD}, together with their nominal values, range and priors.  ``Flat" denotes a prior that is flat within the given range and "Gauss($\mu,\sigma$)" specifies a gaussian prior with mean $\mu$ and variance $\sigma^2$.  The nominal values labeled as "varied" indicate the cases in which these parameters adopt different values during the projection bias studies in this paper (see Table \ref{tab:variables}).}
\begin{ruledtabular}
\begin{tabular}{lccc}
\textrm{Parameter}&
\textrm{Nominal} & \multicolumn{1}{c}{\textrm{Range}}&
\textrm{Prior}\\
\colrule
$\Omega_m$ & 0.267 & (0.10, 0.90) & Flat \\
$A_s \times 10^9$ & 2.870 & (0.50, 5.00) & Flat \\
$n_s$ & varied & (0.87, 1.07) & Flat \\
$\Omega_b$ & varied & (0.03, 0.07) & Flat \\
$h$ & varied & (0.55, 0.91) & Flat \\
$\Omega_\nu h^2 \times 10^3$ & varied & (0.50, 10.0) & Flat \\
$w$ & -1 & fixed & \\
\hline
$b_1$ & 1.42 & (0.8, 3.) & Flat \\
$b_2$ & 1.65 & (0.8, 3.) & Flat \\
$b_3$ & 1.60 & (0.8, 3.) & Flat \\
$b_4$ & 1.92 & (0.8, 3.) & Flat \\
$b_5$ & 2.00 & (0.8, 3.) & Flat \\
\hline
$A_{IA}$ & 0.44 & (-5.0,5.0) & Flat \\
$\eta_{IA}$ & -0.7 & (-5.0,5.0) & Flat \\
\hline
$\Delta z_l^1 \times 10^2$ & \phantom{-}0.8 & (-5.0,5.0) & Gauss(\phantom{-}0.8,0.7) \\
$\Delta z_l^2 \times 10^2$ & -0.5 & (-5.0,5.0) & Gauss(-0.5,0.7) \\
$\Delta z_l^3 \times 10^2$ & \phantom{-}0.6 & (-5.0,5.0) & Gauss(\phantom{-}0.6,0.6) \\
$\Delta z_l^4 \times 10^2$ & \phantom{-}0.0 & (-5.0,5.0) & Gauss(\phantom{-}0.0,1.0) \\
$\Delta z_l^5 \times 10^2$ & \phantom{-}0.0 & (-5.0,5.0) & Gauss(\phantom{-}0.0,1.0) \\
\hline
$\Delta z_s^1 \times 10^2$ & -0.1 & (-10.,10.) & Gauss(-0.1,1.6) \\
$\Delta z_s^2 \times 10^2$ & -1.9 & (-10.,10.) & Gauss(-1.9,1.3) \\
$\Delta z_s^3 \times 10^2$ & \phantom{-}0.9 & (-10.,10.) & Gauss(\phantom{-}0.9,1.1) \\
$\Delta z_s^4 \times 10^2$ & -1.8 & (-10.,10.) & Gauss(-1.8,2.2) \\
\hline
$m^{i (=1,4)} \times 10^2 $ & \phantom{-}1.2 & (-10.,10.) & Gauss(\phantom{-}1.2,2.3) \\
\end{tabular}
\end{ruledtabular}
\end{table}

The main inputs to the calculation of the two-point correlation functions are the number density of galaxies in redshift bins and the power spectrum as a function of wavenumber $k$ and time, or redshift.  The measured galaxy number densities are given by
\begin{equation}
n^i_{g/\kappa}(\chi) = \hat{n}^i_{g/\kappa}(z) \; \frac{d z}{d \chi}
\end{equation}
where $\chi$ is the comoving distance along the line of sight, $z$ is the redshift, $\hat{n}^i_g(z)$ is the normalized galaxy number density for the $i$-th redshift bin, and $\hat{n}^i_\kappa(z)$ is the normalized number density for galaxies that have a shear measurement, also for the $i$-th redshift bin. Throughout the paper, the relation between comoving distance and redshift $z=z(\chi)$ is calculated with the fiducial cosmology ($\Lambda$CDM in our case).  In the DES analysis, the $\hat{n}^i_g(z)$ are referred to as the lens distributions and are given in five redshift bins, and the $\hat{n}^i_\kappa(z)$ are known as the source distributions and are given in four redshift bins.  The uncertainties in the redshift measurements, for both the lens and source galaxies, are included in the analysis by shifting the original measured distribution $\hat{n}^i_{PZ}(z)$ as
\begin{equation}
\hat{n}^i(z) = \hat{n}^i_{PZ}(z-\Delta z^i)
\end{equation}
where the nuisance parameters $\Delta z^i_l$ and $\Delta z^i_s$ for the lenses and sources are allowed to float in the analysis subject to the priors given in Table \ref{tab:parameters}.  With the number density for galaxies with shear measurements, we can calculate the lensing efficiency as
\begin{equation}
\widehat{q}^{\phantom{i}i}(\chi) = \frac{3 \, H_0^2 \, \Omega_m}{2 c^2 \, a(\chi)} \int_\chi^{\chi_h} d\chi^s \, n^i_\kappa(\chi^s) \; \chi \left( 1 - \frac{\chi}{\chi^s} \right)
\label{eq:lensing_efficiency}
\end{equation}
where $H_0$ is the Hubble constant, $c$ the speed of light, $a$ the scale factor, and $\chi_h$ is the comoving distance at the horizon.  Note that, as expected, the lensing efficiency $\widehat{q}^{\phantom{i}i}(\chi)$ is zero for $\chi=0$, or when the source is near the observer, it is small when $\chi$ coincides with the distance to the source galaxies, and it is maximum when $\chi$ is halfway between the observer and the sources.

The lensing efficiency has to be modified to account for the intrinsic alignment (IA) of galaxies \cite{Hirata-Seljak, Hirata-Seljak-erratum}.  The ``non-linear linear alignment" (NLA) model \cite{Bridle-King_IA} was used in the DES Y1 WL analyses for the IA corrections, and Eq. \ref{eq:lensing_efficiency} was modified as \cite{Krause_2017, Troxel_2018}
\begin{equation}
q^i(\chi) = \widehat{q}^{\phantom{i}i}(\chi) - A_{IA} \left( \frac{1+z}{1+z_0} \right)^{\eta_{IA}} \frac{0.0134 \, \Omega_m}{D(\chi)} \; n^i_\kappa(\chi)
\end{equation}
where $z_0=0.62$ and $D(\chi)$ is the linear growth factor,  and the nominal values and priors for the nuisance parameters, $A_{IA}$ and $\eta_{IA}$, are given in Table \ref{tab:parameters}.

In this paper, we use CAMB \cite{CAMB} to calculate the linear power spectrum, and then use this linear power spectrum as input to Halofit \cite{Halofit_1, Halofit_2, Halofit_3} to calculate the nonlinear power spectrum $\widetilde{P}(k,z)$ in bins of $(k,z)$.  The calculation of the power spectrum uses the seven cosmological parameters listed in the first part of Table \ref{tab:parameters}.  The two-point correlation functions are calculated along two different LOS separated by an angle $\theta$.  In the Limber approximation, only the modes in the power spectrum perpendicular to the LOS contribute to the two-point correlations.  When oscillations along the LOS are very long, then $k_\parallel \ll k_\perp$ which leads to $k \approx k_\perp$; and when the oscillations are very short, they average the contribution to the integrals to zero.  Then, in the Limber approximation, only $\widetilde{P}(k_\perp,z)$ enters in the calculation of the two-point correlations with $k_\perp = (l+1/2)/\chi$.  In the flat sky approximation, the two-point correlations are calculated with a double integration involving Bessel functions $J_n(x)$, and the three two-point (3x2pt) correlation functions used in the WL analysis fits are:
\begin{equation}
\widehat{w}^{\phantom{l}i}(\theta) = \int \frac{dl}{2\pi} \; l J_0(l \theta) \int \frac{d\chi}{\chi^2} \; [n^i_g(\chi)]^2 \, \widetilde{P}(k_\perp,z)
\label{eq:gal-clus}
\end{equation}
for the galaxy clustering two-point correlation function,
\begin{equation}
\widehat{\gamma}_t^{\phantom{i}i j}(\theta) = \int \frac{dl}{2\pi} \; l J_2(l \theta) \int \frac{d\chi}{\chi^2} \; n^i_g(\chi) \, q^j(\chi) \, \widetilde{P}(k_\perp,z)
\label{eq:gal-shear}
\end{equation}
for the galaxy-galaxy lensing two-point correlation function, and
\begin{equation}
\widehat{\xi}_{+/-}^{\phantom{i}i j}(\theta) = \!\! \int \frac{dl}{2\pi} \; l J_{0/4}(l \theta) \! \int \frac{d\chi}{\chi^2} \, q^i(\chi) \, q^j(\chi) \,  \widetilde{P}(k_\perp,z)
\label{eq:shear-shear}
\end{equation}
for the cosmic shear two-point correlation function.  In all cases, the indices $i,j$ label the redshift bins of the lens and source $\hat{n}(z)$ distributions.  The two-point correlations calculated above need two corrections.  One is due to the fact that the galaxy number density $n^i_g(\chi)$ only accounts for the visible matter, while light is lensed by all the matter it encounters in its path.  A galaxy bias parameter $b_i$ is added to each lens redshift bin to account for that effect.  DES also accounted for small biases in the galaxy shear measurements by adding a nuisance parameter $m^i$, resulting in the final forms of the two-point correlation functions given by
\begin{align}
w^i(\theta) &= b_i^2 \, \widehat{w}^{\phantom{l}i}(\theta) \\
\gamma_t^{i j}(\theta) &= b_i \, (1+m^j) \, \widehat{\gamma}_t^{\phantom{i}i j}(\theta) \\
\xi_{+/-}^{i j}(\theta) &= (1+m^i) \, (1+m^j) \, \widehat{\xi}_{+/-}^{\phantom{i}i j}(\theta)
\end{align}
The nominal values, intervals, and priors for the galaxy biases $b_i$ and the shear nuisance parameters $m^i$ are given in Table \ref{tab:parameters}.

\subsection{DES Y1 WL Analysis Public Data and Code}\label{subsec:Y1_code}

As mentioned earlier, the data and the code that was used in the weak lensing analysis of the data collected by DES in the first year of running (DES Y1 WL) are publicly available in CosmoSIS \cite{Cosmosis_paper,Cosmosis_program}.  To select the correct versions of the CosmoSIS libraries that were used in the DES Y1 WL analysis, we used the version of CosmoSIS with the date closest to the publication of the DES Y1 WL analysis results \cite{CosmoSIS_branch}.  As described below, some of the parameters in the CosmoSIS version of the DES analysis were changed to better reflect our understanding of how the analysis was done.  This mostly applies to parameter values that are not well documented.

In the DES Y1 WL version stored in CosmoSIS the inner integrals in Equations \ref{eq:gal-clus} to \ref{eq:shear-shear} are  calculated in the range $0.1 \le l \le 5\times 10^5$ with 400 logarithmically spaced points.  After checking that the differences in the analysis between using 200 or 400 points are small (see Appendix \ref{appendix:n_ell}), we decided to reduce the number of points to 200.  Using 400 points would have almost doubled the five months of computing time (using 1728 CPUs) required to do the studies presented in this paper.  It is worth noting that the recently released DES results using three years of data uses 100 points to calculate these integrals.

CosmoSIS offers several possibility for sampling posterior distributions, e.g. Multinest, Polychord, Emcee, etc.  We used Multinest \cite{Multinest_ref} because this is the sampler used in the CosmoSIS version of the DES Y1 WL analysis.  Multinest searches and saves points in increasing values of the likelihood and assigns a weight to each saved point.  In the early iterations the weights are essentially zero because the points are far from the peak of the likelihood, and in late iteration the weights go to zero again because the phase space near the peak of the likelihood goes to zero.  The number of iterations is controlled by the $tolerance$ parameter and we changed its value from $tolerance = 0.1$ to $tolerance \le 10^{-3}$ to make sure that the number of iteration would always be large enough such that the weights would go back to zero after they peaked \footnote{Increasing the value of $tolerance$ from 0.1 to $10^{-3}$ ($10^{-9}$) increases the CPU time by a factor of 1.8 (10.8).}.

As it is standard practice in weak lensing analyses, we processed the output of Multinest using Chainconsumer \cite{Chainconsumer_ref}.  This program uses a Gaussian Kernel Density Estimator algorithm \cite{KDE_algo} to smooth the irregularities produced by the finite number of points provided by posterior samplers like Multinest.  The width of the Gaussian Kernel is controlled by a parameter call KDE.  A marginally low value of KDE=1.0 doesn't produce smooth 95\% credible region edges in two dimensional posteriors and for wide posteriors some times it produces disconnected two dimensional contour, so we selected a value of KDE=1.5.  The consequences of selecting this value are also discussed in Appendix \ref{appendix:KDE_post}.

The first four rows of Table \ref{tab:y1results} show the results of our analysis for different values of the KDE and the $tolerance$ parameters.  The final row shows the results of the DES Y1 WL analysis \cite{DES_Y1_PRD}.  We can see that the differences only amount to a small fraction of the widths of the posteriors and are almost consistent with the inherent fluctuations of samplers like Multinest.
\begin{table}[bht]
\caption{\label{tab:y1results} The first four rows in this table show the results of our analysis of the DES Y1 weak lensing data.  The last row shows the results of the published DES Y1 WL analysis \cite{DES_Y1_PRD}.  We can see that the version of CosmoSIS we will use in our studies reproduces the DES published results very well.}
\begin{ruledtabular}
\begin{tabular}{ccccc}
$\Omega_{m}$& $S_{8}$& $\sigma_{8}$& \textrm{KDE}& \textrm{tol}\\
\colrule\\
$0.267 ^{+0.036}_{-0.021}$ & $0.776 ^{+0.026}_{-0.022}$ & $0.812 ^{+0.054}_{-0.060}$ & 1.0 & 0.1\\
\\
$0.271 ^{+0.033}_{-0.027}$ & $0.776 ^{+0.026}_{-0.024}$ & $0.807 ^{+0.063}_{-0.057}$ & 1.5 & 0.1\\
\\
$0.271 ^{+0.031}_{-0.025}$ & $0.777 ^{+0.024}_{-0.022}$ & $0.808 ^{+0.059}_{-0.052}$ & 1.0 & $10^{-3}$ \\
\\
$0.272 ^{+0.032}_{-0.027}$ & $0.777 ^{+0.026}_{-0.023}$ & $0.810 ^{+0.059}_{-0.056}$ & 1.5 & $10^{-3}$ \\
\\ \hline  \\
 $0.267^{+0.030}_{-0.017}$ & $0.773^{+0.026}_{-0.020}$ & $0.817^{+0.045}_{-0.056}$ & \multicolumn{2}{c}{DES Y1 WL} \\
 \\
\end{tabular}
\end{ruledtabular}
\end{table}
\begin{figure}[hbt]
\includegraphics[width=0.48\textwidth]{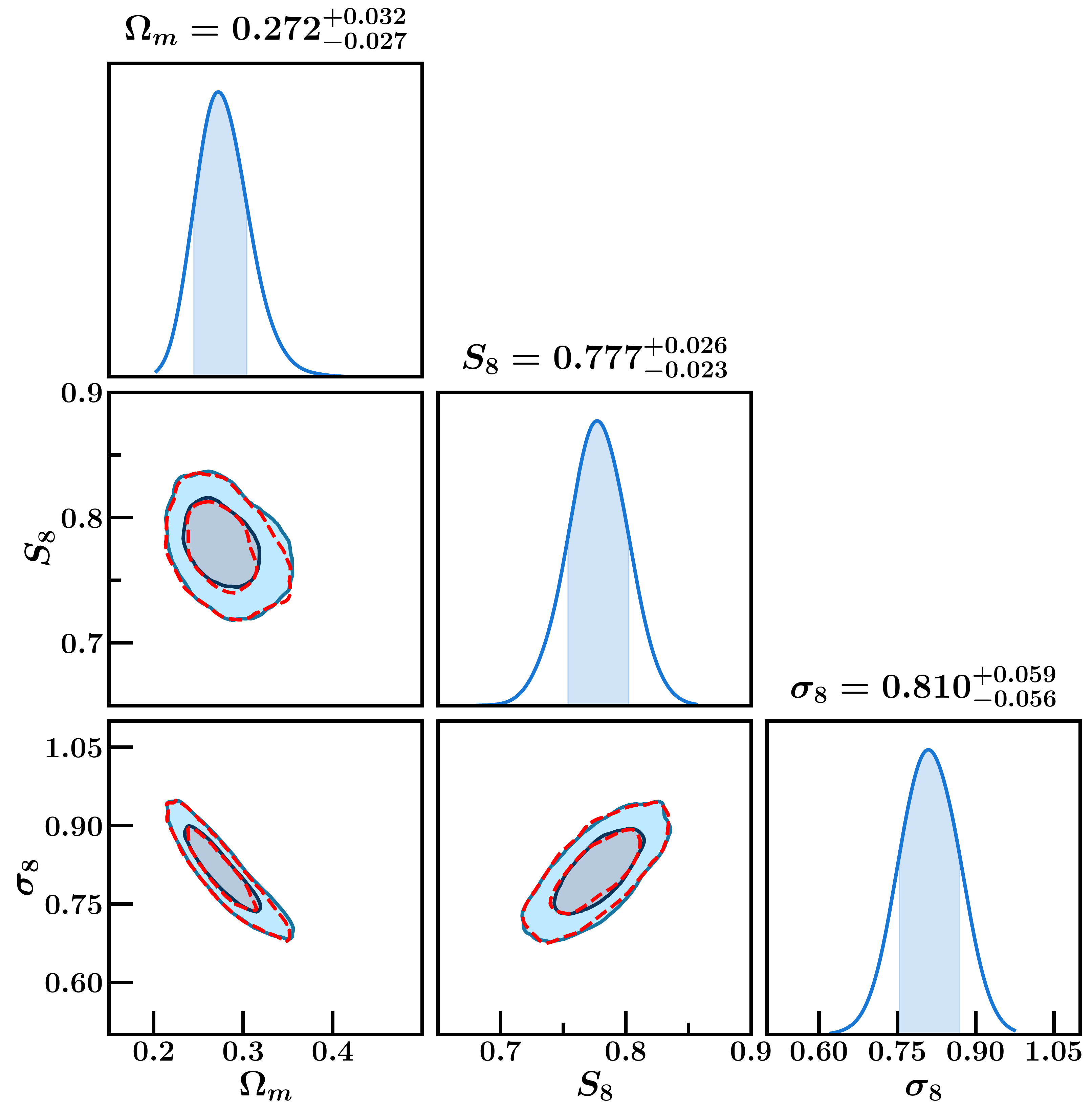}
\caption{The solid (blue) lines in this figure shows the results of our analysis of the DES Y1 WL data using the data and programs publicly available in CosmoSIS.  The plots correspond to row four of Table \ref{tab:y1results}.  The dashed (red) line corresponds to the results of the DES Year 1 3x2pt Weak Lensing analysis (see Fig. 5 in reference \cite{DES_Y1_PRD}).}
\label{fig:Figure5_PRD_us}
\end{figure}

Figure \ref{fig:Figure5_PRD_us} shows one and two dimensional posterior distributions for $\Omega_m$, $\sigma_8$ and $S_8$ corresponding to row four of Table \ref{tab:y1results}.  These plots should be compared with the "DES Y1 all" results in Figure 5 of reference \cite{DES_Y1_PRD}.  These results are shown in Figure \ref{fig:Figure5_PRD_us} with a dashed (red) line.  We can see that the CosmoSIS version we will use in our studies reproduces the published results of the DES Y1 WL analysis very well.

For our studies we need to select a set of nominal or true values for the 26 parameters that enter in the analysis.  The measured values of $\Omega_m$, the biases $b_1$ to $b_5$ and the intrinsic alignment parameters $A_{IA}$ and $\eta_{IA}$ are given in the DES Y1 publication \cite{DES_Y1_PRD}, so we used those values as the true or nominal values of the parameters.  For the nuisance parameters $\Delta z_l^1$ to $\Delta z_l^5$, $\Delta z_s^1$ to $\Delta z_s^4$ and $m^1$ to $m^4$ we used the peak of the priors as our true values.  The measured $\hat{n}_{g/\kappa}(z)$ distributions given by default as input to CosmoSIS assume that $\Delta z^i_{l/s}=0$, so in order to avoid a mismatch between those distributions and the peak of the priors we shifted the $\hat{n}_{g/\kappa}$ distributions to match the priors.  
After shifting the $\hat{n}_{g/\kappa}$ distributions we checked that the analysis of an input data vector calculated using the analysis theory and the true values of all the parameters gives a value of $\chi^2 = 0$ for the fit to the three two-point correlations functions and that the likelihood peak is lined up with the peak of the priors.  This means that the peak of the overall posterior (or MAP point) exactly coincides with the true values of all the parameters.  Therefore any biases observed in the 1D, or 2D, posteriors are entirely due to projection effects.

\begin{table}[bht]
\caption{\label{tab:variables}%
List of nominal or true values used for the cosmological parameters $h$, $\Omega_b$, $n_s$ and $\Omega_\nu h^2$.  The numbers in parenthesis indicate the parameter value as a percent of the parameter's prior range.  All combinations of the values listed in the last three rows ($3^4$ combinations) were studied.  The rest of the true values are given in Table \ref{tab:parameters}.}
\begin{ruledtabular}
\begin{tabular}{cccc}
$h$ & $\Omega_b$ & $n_s$ & $\Omega_\nu h^2 \times 10^3$ \\
\hline
 0.692 & 0.0504 & 0.975 & 0.615 \\
 0.692 & 0.0504 & 0.975 & 4.5 \\
 0.692 & 0.0504 & 0.975 & 9.0 \\
\hline
\multicolumn{4}{c}{\textrm{plus all 81 combinations of the following values}} \\
\hline
 0.64 (25) & 0.04 (25) & 0.92 (25) & 0.615 ($\phantom{1}1.2$) \\
 0.73 (50) & 0.05 (50) & 0.97 (50) & $4.5\phantom{11}$ (42.1) \\
 0.82 (75) & 0.06 (75) & 1.02 (75) & $9.0\phantom{00}$ (89.5) \\
\end{tabular}
\end{ruledtabular}
\end{table}
For the true values of the cosmological parameters with wide likelihood or posterior distributions, $h$, $\Omega_b$, $n_s$ and $\Omega_\nu h^2$, we used the values listed in Table \ref{tab:variables}.  The values of $h$, $\Omega_b$ and $n_s$ of the first three rows in the table correspond to our first determination of those values in the analysis of the DES data.  As we will see in the following section, the projection biases are a function of where the true values sit in relation to the parameter range.  Therefore the previous values were paired with three different values of $\Omega_\nu h^2$: $0.615 \times 10^{-3}$ which corresponds to the minimum allowed by neutrino oscillation experiments, and $4.5 \times 10^{-3}$ and $9.0 \times 10^{-3}$ which sit close to the middle and the upper end of the $\Omega_\nu h^2$ prior range.  For $h$, $\Omega_b$ and $n_s$ we also selected true values of the parameters at 25\%, 50\% and 75\% of their prior's range and paired all their combinations with the $\Omega_\nu h^2$ values described above, giving the $3^4=81$ combinations listed in the last three rows of Table \ref{tab:variables}.  These total of 84 different true values uniformly cover the volume of the four cosmological parameters with wide posteriors.

The posterior distribution for the intrinsic alignment parameter $\eta_{IA}$ is also very wide but its correlations with $\Omega_m$, $\sigma_8$ and $S_8$ are very weak so in our studies we didn't change the true value of this parameter.  Also there are parameters that are strongly correlated, like $\Omega_m$ and the galaxy biases $b_1$ to $b_5$, but all their posteriors fit comfortably within their prior ranges so we didn't change the true values of these parameters either.


\section{Projection Biases}\label{sec:ProjectionBias}

In this section we will describe the results of our projection bias study for the DES Y1 Weak Lensing analysis.  To illustrate the origin of the projection biases, in Subsection \ref{subsec:example} we will describe a simple two dimensional example that contains all the characteristics observed in the real study.  This simple example has several advantages, it is conceptually simple to visualize, it can be easily solved and it runs very fast so accumulating large statistics to reduce fluctuations is not a problem.  In Subsection \ref{subsec:peaks} we will describe the effect that projection biases have in the most likely value of the parameters (the peak of the posteriors) and in Subsection \ref{subsec:intervals} we will describe how projection biases affect the coverage of the 68.27\% credible intervals.  For clarity, the projection bias effects in the most likely values and in the coverage of credible intervals will be presented separately, but they are clearly related.  A shift of the posterior relative to the true value of the parameter will also reduce the number of times the true value falls inside the credible interval. 

\subsection{A Simple Example of Projection Biases}\label{subsec:example}

Given a data set $\mathbf{d}$ and, for a given theory (in our case always $\Lambda$CDM), a set of parameters $\boldsymbol{\theta}$, Bayes theorem states that (see, for example, section ``40-Statistics" in Ref. \cite{Zyla:2020zbs})
\begin{equation}
P(\boldsymbol{\theta}|\mathbf{d}) = \frac{P(\mathbf{d}|\boldsymbol{\theta}) P(\boldsymbol{\theta})}{P(\mathbf{d})}
\label{eq:Bayes_theorem}
\end{equation}
In the DES Y1 WL analysis the vector $\mathbf{d}$ is comprised of the 457 points that are used in the fit of the galaxy clustering, galaxy-galaxy lensing and cosmic shear two-point correlation functions.  $P(\mathbf{d}|\boldsymbol{\theta})$ is the probability density function of obtaining the data set $\mathbf{d}$ given the parameters $\boldsymbol{\theta}$.  When $P(\mathbf{d}|\boldsymbol{\theta})$ is considered as a function of $\boldsymbol{\theta}$ it is called the Likelihood, that is $\mathcal{L}(\boldsymbol{\theta})=P(\mathbf{d}|\boldsymbol{\theta})$.
In WL analysis a $\chi^2$ is calculated using the data $\mathbf{d}$, the theoretical predictions calculated as a function of $\boldsymbol{\theta}$ and the covariance matrix, and the Likelihood is given by $\mathcal{L} \sim e^{-\chi^2/2}$.  
The prior $P(\boldsymbol{\theta})$ includes knowledge external to the analysis and, for all the $\boldsymbol{\theta}$ parameters used in the DES Y1 WL analysis, these priors are listed in Table \ref{tab:parameters}.  The Bayesian evidence $\mathcal{Z} = P(\mathbf{d})$ is given by
\begin{equation}
\mathcal{Z} = \int P(\mathbf{d}|\boldsymbol{\theta}) \, P(\boldsymbol{\theta}) \;  d\boldsymbol{\theta}
\label{eq:evidence}
\end{equation}
The posterior $P(\boldsymbol{\theta}|\mathbf{d})$ is the probability density function for obtaining the parameters $\boldsymbol{\theta}$ when the data $\mathbf{d}$ is given.  
One and two dimensional projections of the weak lensing posteriors, for some of the cosmological parameters involved in our analysis, are shown in Figure \ref{fig:correlations}.  In the two dimensional projections plots, e.g. $h$ vs. $\Omega_m$, we can see clear correlations between parameters whose likelihoods fit well inside the interval, like $\Omega_m, \sigma_8$ and $S_8$ (the last two are not shown in the plots), and parameters for which the likelihood is wider than the specified interval, like $h$, $\Omega_b$, $n_s$ and $\Omega_\nu h^2$.  Projections, or marginalization, over correlated parameters can introduce biases.  This is especially the case when marginalizing over parameters with likelihoods wider than the range of their priors and making measurements of parameters with likelihoods narrower than their priors.  In this section we will illustrate how this works with a simple two dimensional example in which the likelihood for the first and second variable is respectively narrower and wider than their respective priors.  This exemplifies, for example, the 2D-plots of $h$ vs. $\Omega_m$, or $\Omega_b$ vs. $\Omega_m$ shown in Figure \ref{fig:correlations}.  As we will see this simple example has all the same features we will encounter in Sections \ref{subsec:peaks} and \ref{subsec:intervals} when studying the projections biases in the DES Y1 WL analysis.

For our example we will write $\boldsymbol{\theta}=\mathbf{x}=(x,y)$, simplify the posterior notation as $P(\mathbf{x})=P(\boldsymbol{\theta}|\mathbf{d})$ and include the Likelihood peak $\mathbf{\bar{x}}=(\bar{x},\bar{y})$ in the Likelihood notation $\mathcal{L}(\boldsymbol{\theta})=\mathcal{L}(\mathbf{x}|\bar{\mathbf{x}})$.  Then Eq. \ref{eq:Bayes_theorem} in our two dimensional example is written as
\begin{equation}
P(\mathbf{x}) = \frac{1}{\mathcal{Z}} \; \mathcal{L}(\mathbf{x}|\bar{\mathbf{x}}) \, \pi(y)
\label{eq:gen_posterior}
\end{equation}
where the posterior $P(\mathbf{x})$, the likelihood $\mathcal{L}(\mathbf{x}|\bar{\mathbf{x}})$ and the prior $\pi(y)$ are each normalized to one, the prior is given by
\begin{equation}
\pi(y) = \left\{ \begin{array}{cl}
1/(b-a) & \mbox{if } a \le y \le b \\ 0 & \mbox{otherwise} \end{array}\right.
\label{eq:prior_y}
\end{equation}
and the likelihood is given by the two dimensional Gaussian
\begin{equation}
\mathcal{L}(\mathbf{x}|\bar{\mathbf{x}}) = \frac{1}{2\pi \sigma_1 \sigma_2} \; e^{-\frac{1}{2} \chi^2} \;, \mbox{with} \; \chi^2 = \mathbf{u}^T C^{-1} \mathbf{u}
\label{eq:2D-likelihood}
\end{equation}
The parameters $\sigma_1$, $\sigma_2$ and $\alpha$ are the semi-axis and rotation angle of the ellipse with $\chi^2 = 1$, the vector $\mathbf{u}$ is given by  $\mathbf{u} = (u,v) = [(x-\bar{x})/\sigma_x , (y-\bar{y})/\sigma_y]$ with $\sigma_x = [(\cos\alpha / \sigma_1)^2 + (\sin\alpha/\sigma_2)^2]^{-1/2}$ and $\sigma_y$ obtained transposing $\sigma_1 \leftrightarrow \sigma_2$ in $\sigma_x$.  The inverse of the correlation matrix, $C^{-1}$, is given by
\begin{equation}
C^{-1} = \left[ \begin{matrix}
1 & r \\
r & 1 \end{matrix} \right]
\label{eq:2D-correlation}
\end{equation}
with $r=\sigma_x \sigma_y \sin\alpha \cos\alpha \, (1/\sigma_1^2-1/\sigma_2^2)$.

For our studies we selected $\alpha = -20$ degrees, $\sigma_1=1$ and $\sigma_2=5$ which gives posteriors similar to the ones observed in the two dimensional plots involving $\Omega_m$ and displayed in Figure \ref{fig:correlations}.  Since projection biases depend on the relative position of the "True" values with respect to their prior ranges, we selected three symmetric prior intervals and three different true values in each of these intervals.  The selected prior intervals are: $(a,b)=(-y_L,y_L)$, with $y_L =$ 7, 4 and 2, and the selected "True" value of the parameters are: $\mathbf{x_T} = (x_T,y_T) = (0, -R \, y_L)$, with $R=$ 0.9, 0.5 and 0.  We didn't consider values of $R$ greater than one because we assumed that the true value of the parameters are always inside the specified prior range.  Also, we only considered true negative values of $y$ because the corresponding positive values give exactly the same results but with all the distributions reflected along the vertical axis ($x=0$).

\begin{figure}[thb]
\includegraphics[width=0.48\textwidth]{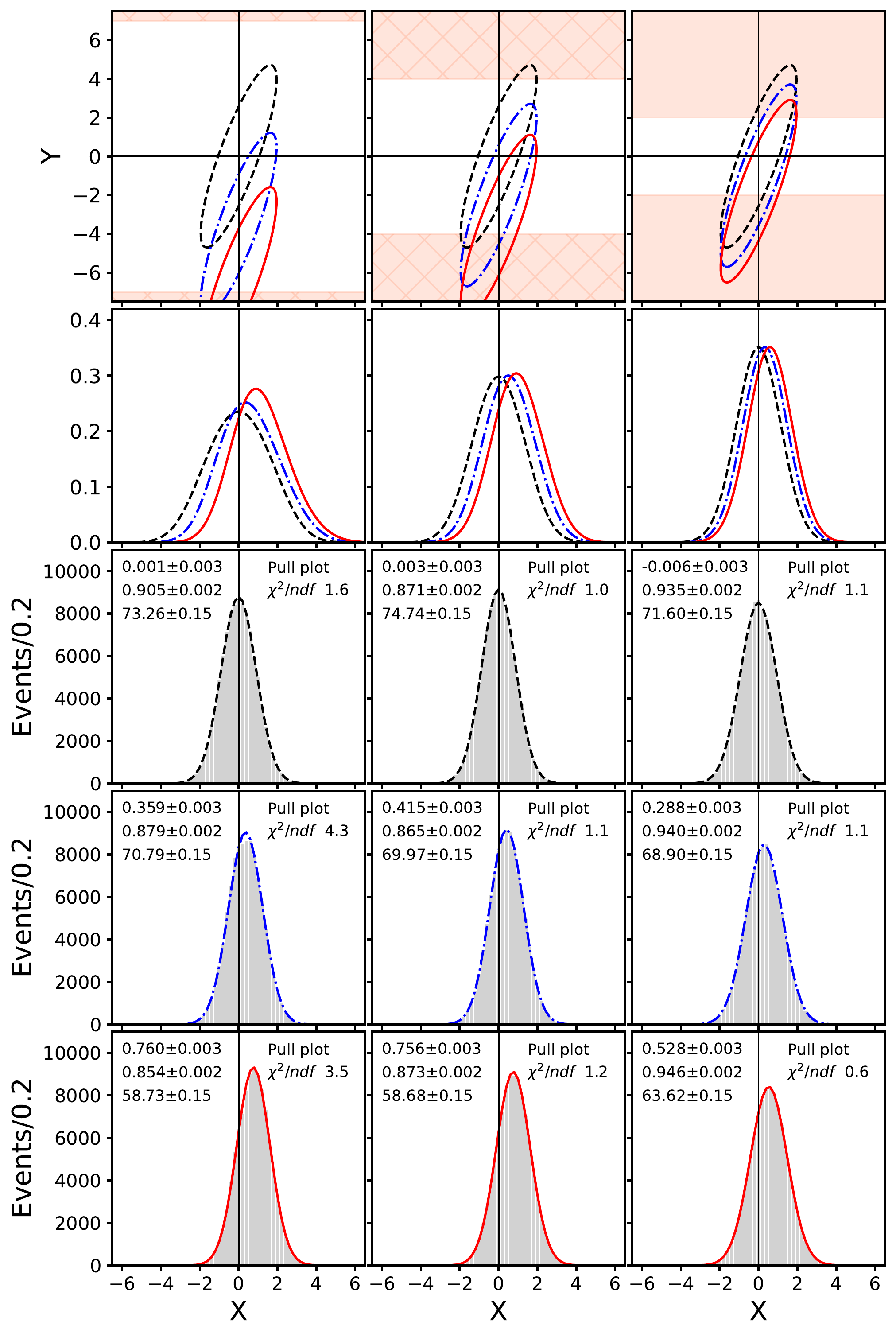}
\caption{Each column in this figure corresponds to different values of the prior $\pi(y)$, with $(a,b)=(-y_L,y_L)$ and $y_L=$ 7, 4 and 2 for the left, center and middle columns.  The shaded area in the top row plots shows the region excluded by the priors.  The different line types correspond to different parameter's true values, $\mathbf{x_T} = (0, -R \, y_L)$ with $R=$ 0, 0.5 and 0.9 for the dashed, dash-dotted and solid lines.  The top row plots show the $\chi^2=1$ contours of $\mathcal{L}(\mathbf{x}|\mathbf{x_T})$ for the values of $\mathbf{x_T}$ given above.  The second row shows the 1D-posteriors when marginalized over $y$, which is to say the projections of the corresponding likelihoods times the priors given in the first row plots.  Rows three to five show the pull histograms, and Gaussian fits, for different true values, with $R=$ 0, 0.5 and 0.9 for rows three, four and five respectively.  See text for more details.}
\label{fig:simple_case}
\end{figure}

The top row of plots in Figure \ref{fig:simple_case} show the $\chi^2=1$ contours of the $\mathcal{L}(\mathbf{x}|\mathbf{x_T})$ likelihood for different values of $R$.  For all plots in Figure \ref{fig:simple_case}, the dashed (black), dash-dotted (blue) and solid (red) lines correspond to $R=$ 0, 0.5 and 0.9.  The shaded areas in the top row plots indicate the values of $y$ excluded by the priors.  Each column in Figure \ref{fig:simple_case} corresponds to different values of the prior, with $y_L=$ 7, 4 and 2 for the left, center and middle columns.

The marginalization of the posterior over the $y$ variable can be calculated exactly as
\begin{eqnarray}
P(x) =&& \frac{1}{\mathcal{Z} (b-a)} \int_a^b dy \; \mathcal{L}(\mathbf{x}|\mathbf{\bar{x}}) \\
=&& \sqrt{ \frac{1-r^2}{2\pi \sigma_x^2}} \; e^{-\frac{1}{2} u^2 (1-r^2)} \left[\frac{\mbox{erf}(t_b) - \mbox{erf}(t_a)}{\mbox{erf}(z_b) - \mbox{erf}(z_a)} \right]
\label{eq:1D-projection}
\end{eqnarray}
where erf$(x)$ is the error function, $t_a = [ v_a + r u]/\sqrt{2}$, $z_a=v_a \sqrt{(1-r^2)/2}$, $v_a=(a-\bar{y})/\sigma_y$, and $t_b$ and $z_b$ are calculated replacing $a$ by $b$ in $v_a$.  The results of this marginalization for $\mathbf{\bar{x}} = \mathbf{x_T}$ is shown in the second row of plots in Figure \ref{fig:simple_case}.  Each dashed, dash-dotted or solid line curve corresponds to the marginalization of the dashed, dash-dotted or solid line two dimensional posteriors shown in the first row plots.  Given that the "True" value is at $x=0$, we can see how the marginalization, or projection, produces biases in the projected or marginalized posteriors, and how this projection biases depend on how the prior is selected relative to the unknown true value of the parameters being marginalized. In the real case these plots would correspond to the one dimensional posteriors of, for example, $\Omega_m$, $\sigma_8$ or $S_8$ when marginalized over all other variables.  From the two top row plots in Figure \ref{fig:simple_case} we can also see that for an ellipse with a negative or clockwise rotation the projection bias will go from maximum positive to minimum negative values when the true value of the wide likelihood parameter goes from the lower to the upper edge of the interval.  The opposite is true if the ellipse has a positive rotation, that is, the bias will go from minimum to maximum values when the true value goes from the lower to the upper edge of the interval.

Biases in the projected distribution would not be a problem if the credible interval corresponding to the 68.27\% area of the posterior included the true value of the parameter 68.27\% of the time.  To study the coverage of the credible intervals we need to simulate doing the same experiment many times and count how often the true value falls inside the 68.27\% interval.  We call this process doing ensemble tests and refer to each simulated experiment as a pseudo-experiment.  If we imagine performing the DES WL experiment many times, subject only to statistical fluctuations, in each new experiment we will get a new value of the data set $\mathbf{d}$ which when fitted will lead to a new likelihood and the extraction of new parameter values $\mathbf{\bar{x}}$.  If the covariance matrix used in the fit was calculated correctly then the values of $\mathbf{\bar{x}}$ will follow a distribution given by $\mathcal{L}(\mathbf{\bar{x}}|\mathbf{x_T)}$ (see Equation \ref{eq:2D-likelihood} for the two dimensional case), where nature has selected and fixed the value of $\mathbf{x_T}$.

Then our ensemble test proceeds according to the following steps.  After selecting the true parameter values $\mathbf{x_T}$, each pseudo-experiment is generated as: 1) calculate a random value of $\mathbf{\bar{x}}$ using the distribution $\mathcal{L}(\mathbf{\bar{x}}|\mathbf{x_T)}$ given in Eq. \ref{eq:2D-likelihood} (notice the different arguments), 2) calculate the 1D-posterior $P(x)$ marginalizing $\mathcal{L}(\mathbf{x}|\mathbf{\bar{x}})$ using Eq. \ref{eq:1D-projection}, 3) calculate the peak and the minimum interval containing the 68.27\% area under $P(x)$, and 4) enter the values in the pull histogram as explained in the next paragraph and check if the true value lies inside the calculated 68.27\% minimum interval.  Repeat, keeping $\mathbf{x_T}$ fixed, until reaching the desired number of pseudo-experiments (one hundred thousand in our case).  The minimum interval $(A,B)$ containing a fixed value of the area under the posterior $P(x)$ is obtained with the condition $P(A)=P(B)$ (see Appendix \ref{appendix:confidence_int}).

Each pseudo-experiment provides the peak of the posterior $x_p$, and two errors $\sigma_-$ and $\sigma_+$ (both positive).  Since in ensemble tests the true value $x_T$ is known we can form the pull distribution as \cite{pull_reference,pull_LucDemortier}: 
\begin{equation}
\left\{ \begin{array}{cl}
\mbox{pull} = (x_p - x_T) / \sigma_+ & \mbox{, if } x_T > x_p \\
\mbox{pull} = (x_p - x_T) / \sigma_- & \mbox{, if } x_T \le x_p \end{array}\right.
\label{eq:pull_definition}
\end{equation}
For asymmetric errors the pull is defined as in the above equation because for $x_T > x_p$ we are comparing $x_T$ with the positive part of the posterior distribution and when $x_T < x_p$ the comparison is with the negative part.

The pull plots are shown in rows three to five of Figure \ref{fig:simple_case}.  Each row of pull plots corresponds to different true values of the parameters $\mathbf{x_T} = (0, -R \, y_L)$, with $R=$ 0, 0.5 and 0.9 for rows three, four and five.  Each column corresponds to different values of the prior $(a,b) = (-y_L,y_L)$, with $y_L=$ 7, 4 and 2 for the left, center and right columns.  The pull histograms were fit to a Gaussian distribution, the mean and $\sigma$ resulting from the fit, with its errors, are shown in the upper and center left legends in each plot.  The lower left legend shows the probability that the 68.27\% interval will include the true value of the parameter.  This value was obtained by counting and its error was calculated assuming binomial statistics.  The legend on the lower right shows the $\chi^2/ndf$ of the fit.

In the absence of biases, and within statistical errors, the mean and rms of the fit to the pull plots should be zero and one and the probability should be 68.27\%.  We verified that this is the case by selecting very large values of $y_L$.  This test also clearly shows that when the prior widths are large in comparison to the widths of the likelihood then all biases disappear, and the pull distributions become Gaussian with mean zero and sigma equals to one.  We also verified that the results are stable with respect to the parameters in the calculation, such as the number of bins used in the fit and the number of point and range in $x$ that were used to calculate the minimum interval containing the 68.27\% area of the posterior.  As a double check we also verified that, within errors, there is agreement in the probability, that the true value will fall inside the credible interval, obtained by counting or by calculating it using the mean and rms of the pulls (see Appendix \ref{appendix:P68_pull}).  The $\chi^2$ probabilities for the fits to the central and right column pull plots of Figure \ref{fig:simple_case} are all larger than 20\%.  For the left column the $\chi^2$ probabilities for the third, fourth and fifth rows are $1.2 \times 10^{-2}$, $1.2 \times 10^{-16}$ and $1.3 \times 10^{-11}$ respectively.  This indicates that in seven out of nine cases the pulls are very Gaussian.

It is clear from the pull plots that even this simple example can exhibit large projection biases depending on the position of the true value relative to the prior range.  We also see that in all cases the width of the pull is smaller than one, which means that the errors calculated from the posteriors are too large.  Based on the rms of the pull plots, we calculate that the credible intervals are between 5\% and 15\% larger than they should be.  Also even with these inflated intervals, we can see from the pull plots that the coverage can be as low as 58.7\% due to the bias in the position of the posteriors.  In the next two sections we will observe the same biases in the DES Y1 Weak Lensing analysis.  We will also see that these biases can be much larger than the ones encountered in this section, the reason for this is that in the real analysis there are four correlated wide distributions that contribute to the projection biases: $h$, $\Omega_b$, $n_s$ and $\Omega_\nu h^2$.  An example of these correlations is shown in Figure \ref{fig:correlations}.

\begin{figure}[bht]
\includegraphics[width=0.48\textwidth]{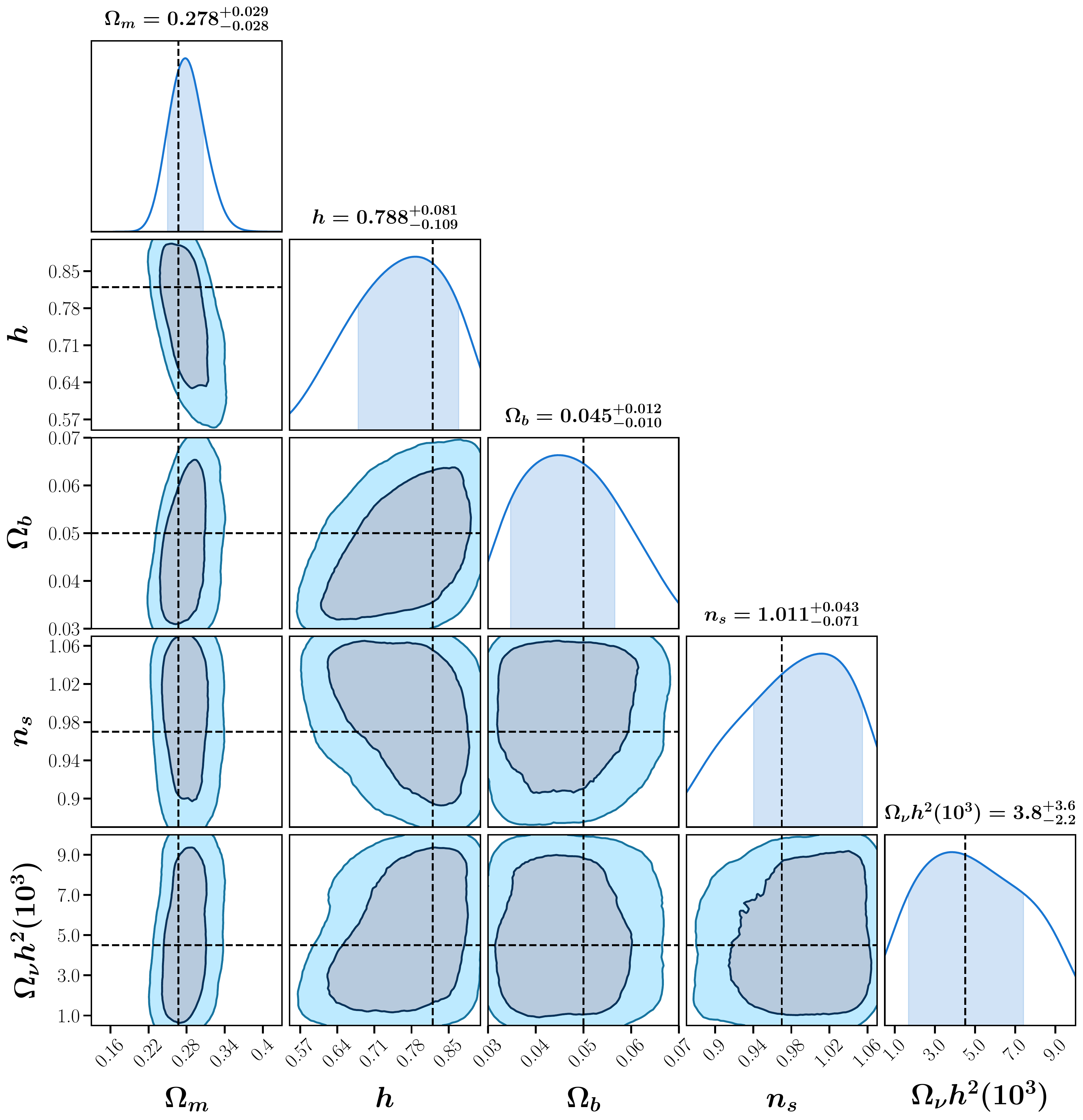}
\caption{Selected one and two dimensional posterior plots from the analysis of a synthetic data vector with true parameter values $h=0.82$, $\Omega_b=0.05$, $n_s=0.97$ and $\Omega_\nu h^2 = 4.5 \times 10^{-3}$.  The true or nominal values for the rest of the parameters are given in Table \ref{tab:parameters}.}
\label{fig:correlations}
\end{figure}

\subsection{Biases in the Posterior Peaks}\label{subsec:peaks}

To study coverage of confidence intervals one imagines doing the same experiment a very large, or infinite, number of times.  Then, in the absence of systematic errors the true value of the quantity we are trying to measure will lie inside the 68.27\% confidence interval 68.27\% of the time, and the proper average of the most likely values of that quantity will converge to its true value.  In our case we simulated this process by fluctuating, using the covariance matrix, an exact calculation of the three two-point correlation functions and then used this fluctuated two-point correlations as input data to the analysis.  This process, which as pointed out in Section \ref{sec:introduction} we are calling doing ensemble tests, is very CPU time intensive so for most of our studies we used a "synthetic data vector".  We will show that for studying biases in the position of the posterior distributions there is good agreement between the two methods.

A synthetic data vector is composed of three two-point correlation functions, as a function of the angles used in the analysis, calculated using the true or nominal values of the parameters listed in Table \ref{tab:variables} and the formalism summarized in Section \ref{subsec:Y1_model}.  This synthetic data vector is then used to replace the input data in the DES Y1 WL analysis leaving everything else unchanged, e.g. the $\hat{n}_{g/\kappa}(z)$ distributions, covariance matrix, priors, etc.  In this case the theory used in the analysis matches exactly the theory that generated the input data which eliminates all systematic errors, for example, from measurement errors or coming from discrepancies between the analysis theory and the data.  Furthermore, since the true parameter values are known, this procedure allows for the study of sources of biases in the analysis that are not related to systematic errors.  In this section we will discuss the results of the analysis of the 84 synthetic data vectors generated with the true values listed in Table \ref{tab:variables}.  The ensemble tests performed on a selection of five of the 84 synthetic data vectors will be discussed in the following section.

Figure \ref{fig:correlations} shows results of the analysis of a synthetic data vector generated with true values $h=0.82$, $\Omega_b=0.05$, $n_s=0.97$ and $\Omega_\nu h^2 = 4.5 \times 10^{-3}$, corresponding to fractions of the parameter range of 0.75, 0.5, 0.5 and 0.421.  All the other true or nominal values of the parameters are listed in Table \ref{tab:parameters}.
To understand the behaviour of projection biases it is worth paying attention to the correlations seen in Figure \ref{fig:correlations} between $\Omega_m$, and the parameters with wide posteriors $h$, $\Omega_b$, $n_s$ and $\Omega_\nu h^2$.  Going down the first column, we see that there is a positive, or counterclockwise, correlation with $h$, and then the correlation alternates sign for $\Omega_b$, $n_s$ and $\Omega_\nu h^2$.  This pattern of correlations is typical and fairly independent of the true values of the parameters, except that they mostly reverse sign for $\sigma_8$ and $S_8$.

In the example given in Section \ref{subsec:example} we saw that the projection bias becomes positive and maximum when there is a negative, or clockwise, correlation and the true value of the parameter with the wide likelihood, or wide posterior, is at the lower end of its range.  For a positive, or counterclockwise, correlation the opposite is true, the projection bias is at its maximum positive value when the true value of the parameter with the wide distribution is at the upper end of its range.  For more than one parameter with wide posteriors the contributions to the projection bias will add, then we should expect that the bias in $\Omega_m$ will have a maximum positive value when the true value of $h$ is near the upper end of the range, and the true values of $\Omega_b$, $n_s$ and $\Omega_\nu h^2$ are respectively near the lower, upper and lower ends of their ranges.  And this is exactly what we will observe.  Also we would expect that the bias reaches it lowest, or negative, value when the above four parameters are at the opposite end of their ranges, and again this is what we will observe.

\begin{figure}[hbt]
\includegraphics[width=0.48\textwidth]{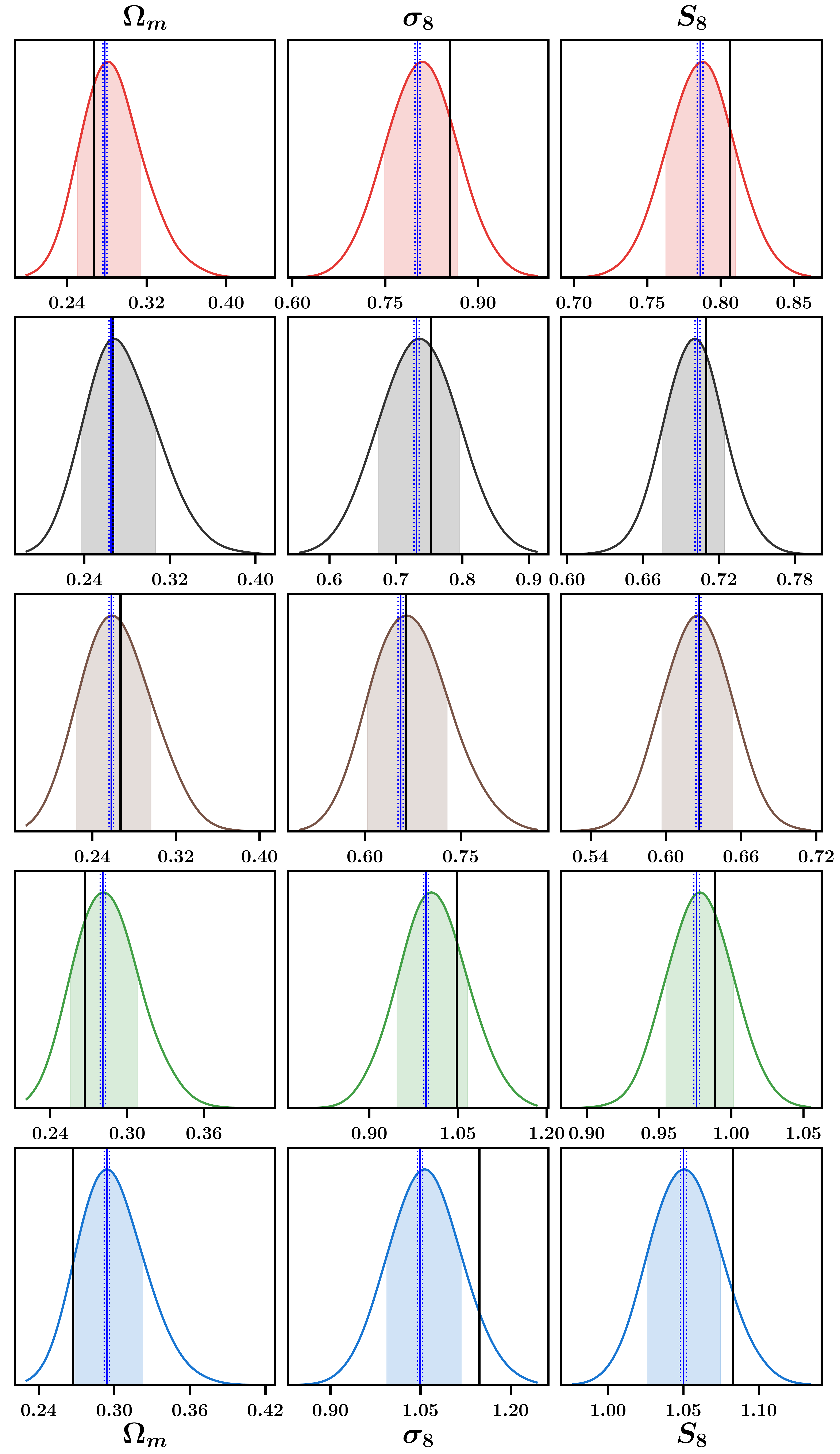}
\caption{Posterior distributions from the analysis of synthetic data vectors created with the true parameter values listed in Tables \ref{tab:ensembles} and \ref{tab:parameters}.  The columns correspond to $\Omega_m$, $\sigma_8$ and $S_8$.  The rows correspond to different values on the true parameters in the same vertical order they are listed in Table \ref{tab:ensembles}.  The vertical solid lines correspond to the true value of the parameters, the vertical solid lines flanked by dotted lines correspond to the ensemble tests results.}
\label{fig:posteriors}
\end{figure}

Figure \ref{fig:posteriors} shows the $\Omega_m$, $\sigma_8$ and $S_8$ posterior distributions for the five cases in which we have results for both the synthetic data vector analysis and ensemble tests.  The true values of $h$, $\Omega_b$, $n_s$ and $\Omega_\nu h^2$ used to calculate the plots in the figure are listed in Table \ref{tab:ensembles}, the rest of the true parameter values are given in Table \ref{tab:parameters}.
\begin{table}[bht]
\caption{\label{tab:ensembles}%
This table lists the nominal or true values of the cosmological parameters $h$, $\Omega_b$, $n_s$ and $\Omega_\nu h^2$ for which the projections biases were studied using both ensemble tests and analyzing a single synthetic data vector.  The numbers in parenthesis indicate the parameter value as a percent of the parameter's prior range.  The rest of the true values are given in Table \ref{tab:parameters}.}
\begin{ruledtabular}
\begin{tabular}{c|cccc}
 \# & $h$ & $\Omega_b$ & $n_s$ & $\Omega_\nu h^2 \times 10^3$ \\
\hline
 1 & 0.692 (39) & 0.0504 (51) & 0.975 (53) & 0.615 ($\phantom{1}1.2$) \\
 2 & 0.692 (39) & 0.0504 (51) & 0.975 (53) & $4.5\phantom{11}$ (42.1) \\
 3 & 0.692 (39) & 0.0504 (51) & 0.975 (53) & $9.0\phantom{00}$ (89.5) \\
 4 & $0.82\phantom{0}$ (75) & $0.04\phantom{00}$ (25) & $1.02\phantom{0}$ (75) & $4.5\phantom{11}$ (42.1) \\
 5 & $0.82\phantom{0}$ (75) & $0.04\phantom{00}$ (25) & $1.02\phantom{0}$ (75) & 0.615 ($\phantom{1}1.2$) \\
\end{tabular}
\end{ruledtabular}
\end{table}
The posteriors for the different cases in Figure \ref{fig:posteriors} are given in the same vertical order as in Table \ref{tab:ensembles}.  From left to right the columns show the posteriors for $\Omega_m$, $\sigma_8$ and $S_8$, and the shaded areas show the 68\% area of the posteriors.  The vertical solid lines show the true values of the corresponding parameters.  We can clearly see that in most cases the true values and the peaks of the posteriors do not agree, which is the effect of the projection biases.  We can also clearly see that these biases are a function of the true parameter values, with the biases being smaller when the true value of the parameters sit close to the middle of the prior interval.  This is the same effect observed in the simple example studied in Section \ref{subsec:example} and exemplified in the second row plots of Figure \ref{fig:simple_case} in that section.

The DES collaboration recently released the weak lensing results of their analysis of the first three years of data.  The results from the fit to all three two-point correlation functions (3x2pt analysis) are given in Ref. \cite{DES_Y3_3x2pt}.  The validation of the theoretical model used in the analysis is given in Ref. \cite{DES_Y3_3x2pt_validation}.  Figure 2 of this reference shows the projection biases for the true values of $h=0.69$, $\Omega_b=0.048$, $n_s=0.97$ and $\Omega_\nu h^2 = 0.83 \times 10^{-3}$.  As in our case, we can see that the biases for $\Omega_m$ and $S_8$ approach the 1 $\sigma$ level, but these biases were studied for only one set of true values of the parameters.

The vertical solid lines flanked by dotted lines in Figure \ref{fig:posteriors} show the results of the ensemble tests.  The solid lines show the most likely value obtained by multiplying the posteriors of all 220 runs in each ensemble test, the flanking dotted lines show the plus and minus one sigma errors obtained in the same way.  We can see that the errors in the most likely value of the ensemble tests are very small and that there is very good agreement between the peak of the posteriors calculated using a synthetic data vector and the ensemble tests.  This gives us confidence that we can discuss projection biases in the peak of the posteriors using mostly synthetic data vectors.

\begin{figure*}[htb]
\includegraphics*[width=0.7\textwidth]{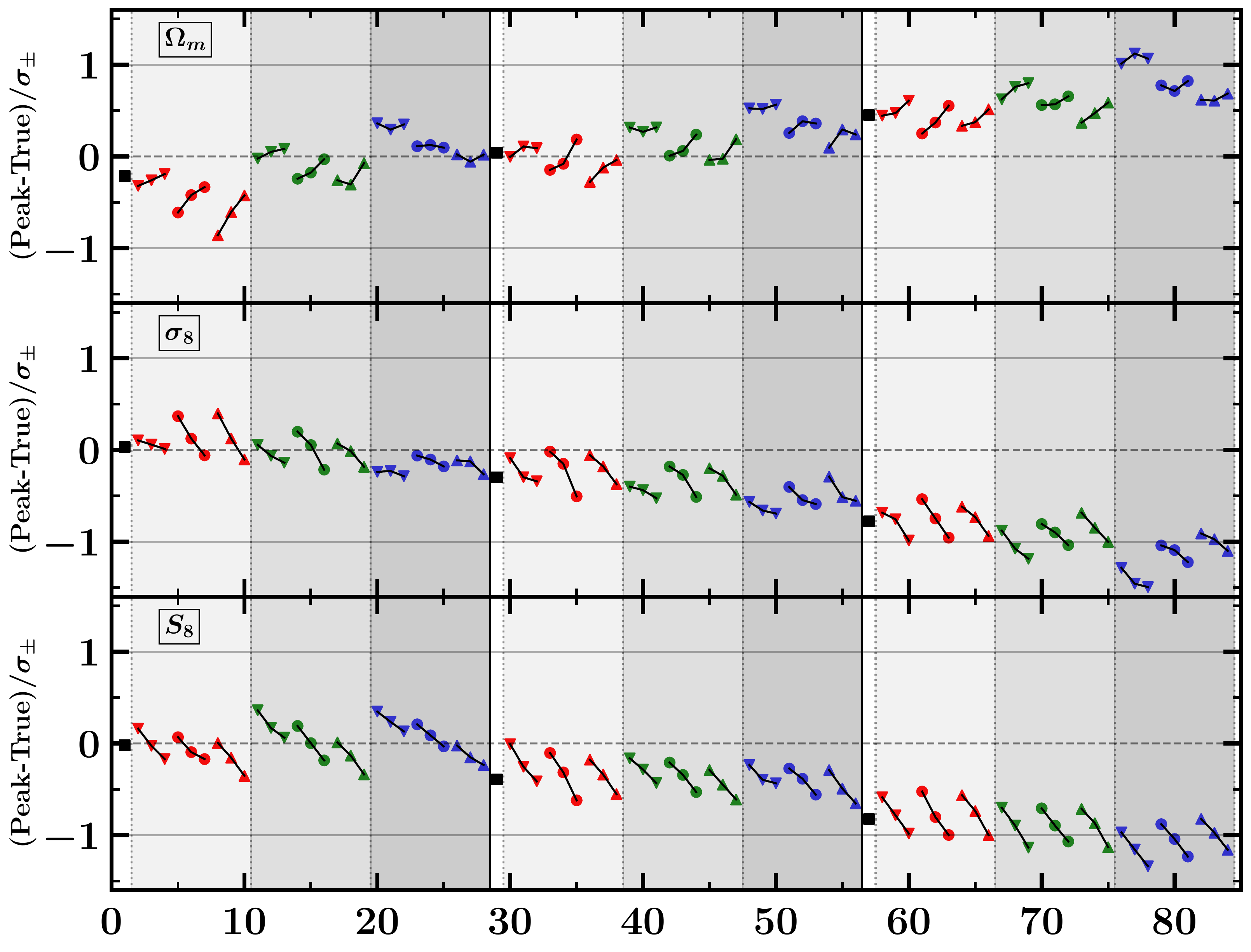}
\caption{Projection bias plots for the 84 combinations of parameter true values given in Tables \ref{tab:parameters} and \ref{tab:variables}.  The vertical scales show the peak of the posterior minus the true value divided by $\sigma_+$ ($\sigma_-$) if the peak of the posterior is smaller (greater) than the true value.  The horizontal scale is a case number from 1 to 84.   The groups of three points joined by a line correspond to different values of $n_s$.  The downward triangles, circles and upward triangles correspond to different values of $\Omega_b$.  The shades of light, medium and darker grey correspond to different values of $h$.  And the large left, middle and right groups of 27 points correspond to different values $\Omega_\nu h^2$.  The values of $n_s$, $\Omega_b$ and $h$ increase from left to right taking the values of 25\%, 50\% and 75\% of the prior's range.  $\Omega_\nu h^2$ decrease from left to right with values of 89.5\%, 42.1\% and 1.2\% of the prior's range.  The three solid squares correspond to the top three cases listed in Table \ref{tab:variables}.}
\label{fig:peak_bias}
\end{figure*}

Figure \ref{fig:peak_bias} shows the bias in the peak of the posteriors for all 84 combinations of true values listed in Table \ref{tab:variables}, five of these entries correspond to the peak of the posteriors shown in Figure \ref{fig:posteriors}.  In all cases the posteriors were calculated using synthetic data vectors.  The vertical scales show the pull values, that is the peak of the posterior minus the true value divided by $\sigma_+$ ($\sigma_-$) if the peak of the posterior is smaller (greater) than the true value.  The selection of which $\sigma$ to use in the denominator is the same as in the definition of the pull in Eq. \ref{eq:pull_definition}, when the peak of the posterior is smaller than the true value we are comparing the true value with the positive side of the posterior and when the peak of the posterior is larger than the true value the comparison is with the negative part of the posterior.  The horizontal scale is a case number from 1 to 84.  Each group of three points joined by a line correspond to three different values of $n_s$, and from left to right these values are 25\%, 50\% and 75\% of the interval prior.  The downward pointing triangles, the circles and the upward pointing triangles correspond to values of $\Omega_b$ of 25\%, 50\% and 75\% of the prior's interval.  The groups with background shades of light, medium and darker grey correspond to values of $h$ of 25\%, 50\% and 75\% of the prior's interval respectively.  The large left, middle and right groups of 27 points correspond to values $\Omega_\nu h^2$ of 89.5\%, 42.1\% and 1.2\% of the prior's interval respectively.  Note that while the true values of $n_s$, $\Omega_b$ and $h$ increase from left to right for $\Omega_\nu h^2$ the true values increase from right to left.  Finally the three solid squares in between the larger groups of 27 points correspond to the top three cases listed in Table \ref{tab:variables}.

As discussed earlier in this section, the overall pattern of the projection biases in Figure \ref{fig:peak_bias} can be understood with the correlations observed in the first column of Figure \ref{fig:correlations} and the simple example given in Section \ref{subsec:example}.  In the top row plot of Figure \ref{fig:peak_bias} we see that the overall projection biases for $\Omega_m$ move towards increasing positive values with increasing values of $n_s$ and $h$, and decreasing values of $\Omega_b$ and $\Omega_\nu h^2$.  At the same time from the first column in Figure \ref{fig:correlations} we can see that the correlations with $\Omega_m$ are positive (ellipse rotated counterclockwise) with $n_s$ and $h$, and negative for $\Omega_b$ and $\Omega_\nu h^2$.  In our simple example we saw that for positive (negative) correlations the projection bias moves towards more positive values when the value of the parameter increases (decreases).  So the projection biases for $\Omega_m$ behave as expected.  Since $\sigma_8$ is anti-correlated with $\Omega_m$ we would expect the opposite behaviour for $\sigma_8$ and this is what we observe comparing the first and second row plots in Figure \ref{fig:peak_bias}.  The behaviour of the projection bias for $S_8$ follows closely that of $\sigma_8$ except for the first light grey area where the very low values of the bias for $\Omega_m$ pull down the bias for $S_8$.

Not surprisingly then, the largest projection biases in our 84 cases occurs when the true values of $n_s$, $\Omega_b$, $h$ and $\Omega_\nu h^2$ are at 75\%, 25\%, 75\% and 1.2\% of their respective ranges (seventh points from right to left).  The biases in this case are close to 1 $\sigma$ for $\Omega_m$ and to -1.5 $\sigma$ for $\sigma_8$ and $S_8$.  The projection biases would actually be larger if the true values would be pushed closer to the limits of their ranges, but we have assumed that the parameter ranges are selected conservatively and for $n_s$, $\Omega_b$ and $h$ stayed within 25\% of the edges of the parameter ranges.  Of course for $\Omega_\nu h^2$ we selected true values very close to the lower range limit because a very low neutrino mass first eigenstate is always a real possibility.  As previously discussed the opposite values of the projection biases for $\Omega_m$ and $\sigma_8$ occur at the opposite true values of $n_s$, $\Omega_b$, $h$ and $\Omega_\nu h^2$ that is at 25\%, 75\%, 25\% and 89.5\% of their respective ranges (eighth points from left to right).  In this case the biases are about -1 $\sigma$ for $\Omega_m$ and 0.5 $\sigma$ for $\sigma_8$.  The highest positive bias for $S_8$ is also about 0.5 $\sigma$ but it is shifted due to the influence of the $\Omega_m$ biases on $S_8$ in the first light grey band.

\subsection{Biases in the 68\% Credible Intervals}\label{subsec:intervals}

\begin{figure}[hbt]
\includegraphics[width=0.48\textwidth]{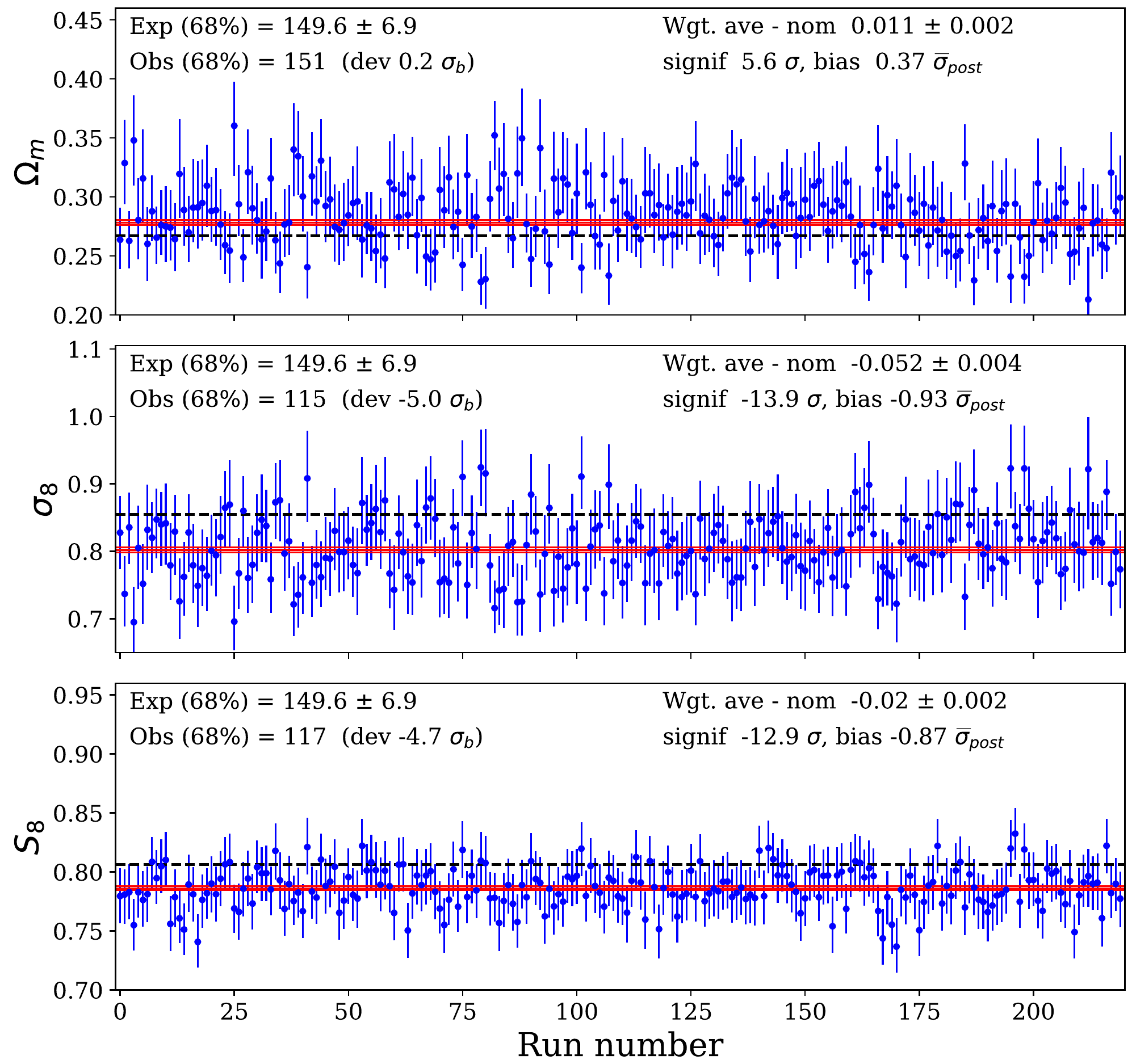}
\caption{Ensemble test results for $\Omega_m$, $\sigma_8$ and $S_8$.  Each run is represented by a dot with error bars displaying the peak of the posterior and its errors.  The horizontal axis displays the run number in the ensemble test.  The dashed horizontal lines show the true values (see first row in Table \ref{tab:ensembles}) and the three horizontal solid (red) lines show the weighted average and one sigma errors for the combination of all 220 runs.}
\label{fig:Case1_ensembles}
\end{figure}

Five of the 84 cases discussed in the previous section were selected for ensemble tests.  The true value of the parameters for these five cases are listed in Table \ref{tab:ensembles}.  For each of these cases, 220 input data vectors were generated fluctuating the synthetic data vector using the covariance matrix.  These new data vectors were then processed using the regular DES Y1 WL analysis.  Results for $\Omega_m$, $\sigma_8$ and $S_8$ of the 220 runs corresponding to the first row of true values in Table \ref{tab:ensembles} are shown in Figure \ref{fig:Case1_ensembles}.
Each dot with error bars in this figure represents the peak of the posterior and its errors for all 220 runs in this ensemble test.  The dashed line represents the true value for the corresponding parameter.  The weighted average ($\bar{x}_W$) and the error ($\sigma$) for the combined runs were obtained by multiplying the posteriors for all 220 runs.  This weighted average minus the true or nominal value and the one sigma error are shown in the upper right side of the plots.  These values were also entered as the vertical line flanked by dotted lines in the upper row plots of Figure \ref{fig:posteriors}; the vertical line corresponds to $\bar{x}_W$ and the flanking dotted lines to $\bar{x}_W \pm \sigma$.  The vertical lines with flanked dotted lines in the other fours rows of the same figure were calculated in a similar way from their corresponding ensemble tests.  We can see that there is consistency between the biases in the peak of the posteriors obtained using a synthetic data vector and those obtained from ensemble tests.  Also shown in the upper right side of the plots in Figure \ref{fig:Case1_ensembles} are the significance defined as ($\bar{x}_W$ - true)/$\sigma$ and the bias defined as ($\bar{x}_W$ - true)/$\bar{\sigma}_{post}$, where $\bar{\sigma}_{post} = (\sigma^+_{post}+\sigma^-_{post})/2$ is the average of the positive and negative errors of the posterior.  We can see that 220 runs are enough to determine the biases with a high degree of significance.  The upper left side of each plot shows the expected (Exp) and observed (Obs) number of times the true value falls inside the 68.27\% credible interval of the posterior.  The error in the expected value ($\sigma_b$) was calculated using binomial statistics.  The deviation is defined as the observed minus the expected number of events divided $\sigma_b$, dev = (Obs-Exp)/$\sigma_b$.  Given the biases in the posteriors it is not surprising that for $\sigma_8$ and $S_8$ the number of observed cases where the true value falls inside the posterior's 68.27\% credible interval is substantially lower than expected with a high degree of significance.  

\begin{figure*}[hbt]
\includegraphics[width=0.63\textwidth]{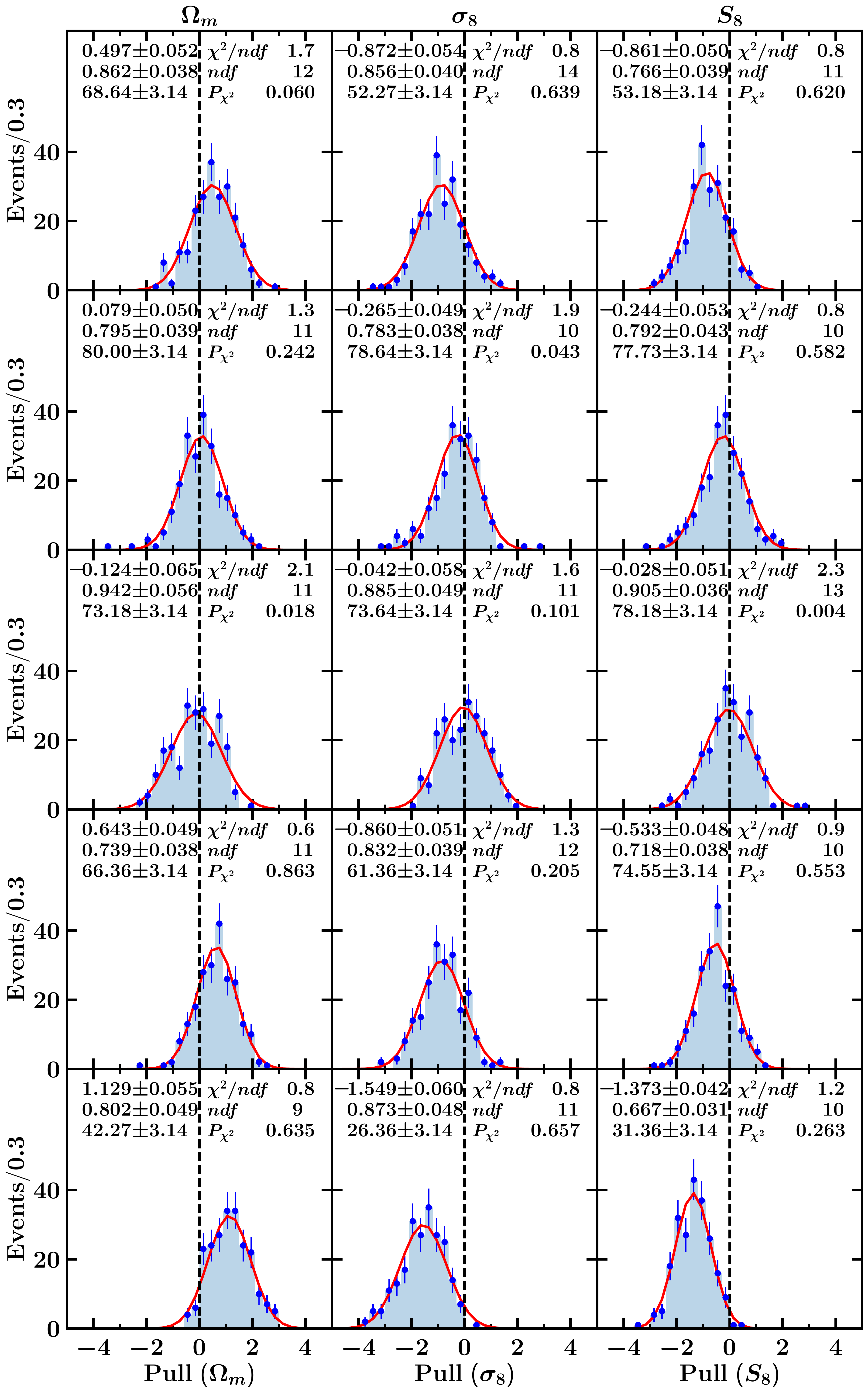}
\caption{Histograms of the pull values shown in Equation \ref{eq:pull_definition} for the five ensemble tests with the true values given in Table \ref{tab:ensembles}.  The vertical order is the same as in the table, the points with error bars show the number of entries in each bin with their binomial errors.  From top to bottom the left side legends show the mean ($\bar{x}_p$) and sigma ($\sigma_p$) of the Gaussian fit to the histogram and the probability ($P_{68}$) that the true value will lie inside the 68.27\% credible interval.  The right legends show $\chi^2/ndf$, the number of degrees of freedom ($ndf$) and the $\chi^2$ probability of the Gaussian fits.  }
\label{fig:pull_plot}
\end{figure*}

The best way to visualize the projection biases in both the most likely values and the credible interval of the posteriors is with pull plots.  The pull is defined as in Equation \ref{eq:pull_definition}.  Figure \ref{fig:pull_plot} shows the pull plots for the five ensemble test cases outlined in Table \ref{tab:ensembles}.  The vertical order in the pull plots is the same as in the table.  
For infinity statistical samples and in the absence of biases these plots will be Gaussian distributions with mean $\bar{x}_{p}=0$ and sigma equals to $\sigma_{p}=1$.  The solid lines in the plots show the Gaussian fit to the histograms.  The interval used in the fit extends to $\pm 3.0 \sigma$ from the peak of the distributions.  We checked that the fit results are stable with the selection of this interval and with the selection of the bin size.  From top to bottom the legends on the right show $\chi^2/ndf$, the number of degrees of freedom ($ndf$) and the $\chi^2$ probability for the fit.  The value of $ndf+3$ is equal to the number of bins used in the fit, which changes for every plot because of the $\pm 3.0 \sigma$ selection criteria.  The top and center legends on the left side of each plot show the average ($\bar{x}_p$) and sigma ($\sigma_p$) of the Gaussian fits.  The probability that the true value will be inside the 68.27\% credible interval ($P_{68}$) is given by the lower value on the left legends.  $P_{68}$ was calculated by simply counting the number of times the true value was inside the 68.27\% area of the posterior.  We can see that overall the fits look very Gaussian, but there are large biases in both the mean and sigma of the fits.  The values of $\sigma_p$ range from 0.67 to 0.94 indicating that in all cases the posterior credible intervals are too wide (the average width of the CI is $1/\sigma_{p}$).  This is consistent with the pulls observed in our simple example in Section \ref{subsec:example} and in Figure \ref{fig:simple_case}.  In many cases we can also see a substantial shift in the pull distribution to the point that even with very inflated credible intervals the $P_{68}$ probability can be as low as 26.4\%.

We can check the consistency of the pulls by calculating the $P_{68}$ probability from the pull parameters using the expression (see Appendix \ref{appendix:P68_pull})
\begin{equation}
P_{68}(\mbox{pull})= \frac{1}{2} \left\{ \mbox{erf}\left(  \frac{1-\bar{x}_p}{\sqrt{2} \; \sigma_p} \right) + \mbox{erf}\left(  \frac{1+\bar{x}_p}{\sqrt{2} \; \sigma_p} \right)  \right\}
\label{eq:p68_pull}
\end{equation}
and then comparing it with the $P_{68}$ probability obtained by counting.  The $P_{68}$ probability obtained by counting is shown in the first row in each of the $\Omega_m$, $\sigma_8$ and $S_8$ groups in Table \ref{tab:p68_check}.  From left to right each column in the table corresponds to the top to bottom plots in Figure \ref{fig:pull_plot}.  The second row in each group corresponds to the $P_{68}$ probabilities calculated from the pull parameters using Equation \ref{eq:p68_pull}, and the third row is the difference between the two calculations divided by the error in the $P_{68}$ probability obtained by counting.  We can see that the two ways of obtaining $P_{68}$ are completely consistent with each other.
\begin{table}[bht]
\caption{\label{tab:p68_check}%
Table of probabilities that the true value will fall inside the 68.27\% interval.  For each parameter the first row shows the values obtained by counting, the second row are the values calculated from the pull plots (see Equation \ref{eq:p68_pull}) and the third row is the difference divided by the error in the probability obtained by counting.}
\begin{ruledtabular}
\begin{tabular}{c|ccccc}
  & 1 & 2 & 3 & 4 & 5 \\
\hline
 $\Omega_m$ & 68.6$\pm$3.1 & 80.0$\pm$3.1 & 73.2$\pm$3.1 & 66.4$\pm$3.1 & 42.3$\pm$3.1 \\
   & 67.9$\pm$2.1 & 78.9$\pm$2.2 & 70.7$\pm$2.9 & 67.2$\pm$2.5 & 43.2$\pm$2.6 \\
   & -0.2 & -0.3 & -0.8 & 0.3 & 0.3 \\
\hline
 $\sigma_8$ & 52.3$\pm$3.1 & 78.6$\pm$3.1 & 73.6$\pm$3.1 & 61.4$\pm$3.1 & 26.4$\pm$3.1 \\
  & 54.5$\pm$2.3 & 77.3$\pm$2.2 & 74.1$\pm$2.6 & 55.4$\pm$2.3 & 26.3$\pm$2.4 \\
  & 0.7 & -0.4 & 0.1 & -1.9 & 0.0 \\
\hline
 $S_8$ & 53.2$\pm$3.1 & 77.7$\pm$3.1 & 78.2$\pm$3.1 & 74.5$\pm$3.1 & 31.4$\pm$3.1 \\
  & 56.4$\pm$2.5 & 77.2$\pm$2.5 & 73.1$\pm$1.9 & 72.6$\pm$2.5 & 28.8$\pm$2.3 \\
  & 1.0 & -0.2 & -1.6 & -0.6 & -0.8 \\
\end{tabular}
\end{ruledtabular}
\end{table}

As expected from our simple example in Section \ref{subsec:example} the results in this section clearly show that the posterior credible intervals can be substantially inflated.  The question we may ask here is, are there contributions to the inflation of errors other than projection biases?  Another source of error inflation is the Gaussian Kernel Density Estimator (KDE) algorithm \cite{KDE_algo} used by programs like Chainconsumer \cite{Chainconsumer_ref}.  This program is widely used in weak lensing analysis to process the output of samplers like Multinest, Polychord or Markov chain Monte Carlo programs, and it was used in this publication to process the output of Multinest (see for example the results in Table \ref{tab:y1results}).  In the Gaussian KDE algorithm each point is replaced by a finite width Gaussian distribution which can artificially widen the posteriors and give as a consequence inflated credible intervals.  In Appendix \ref{appendix:KDE_post} we show that this inflation of errors is small and cannot account for the larger credible intervals observed in this Section.

\begin{figure}[hbt]
\includegraphics[width=0.48\textwidth]{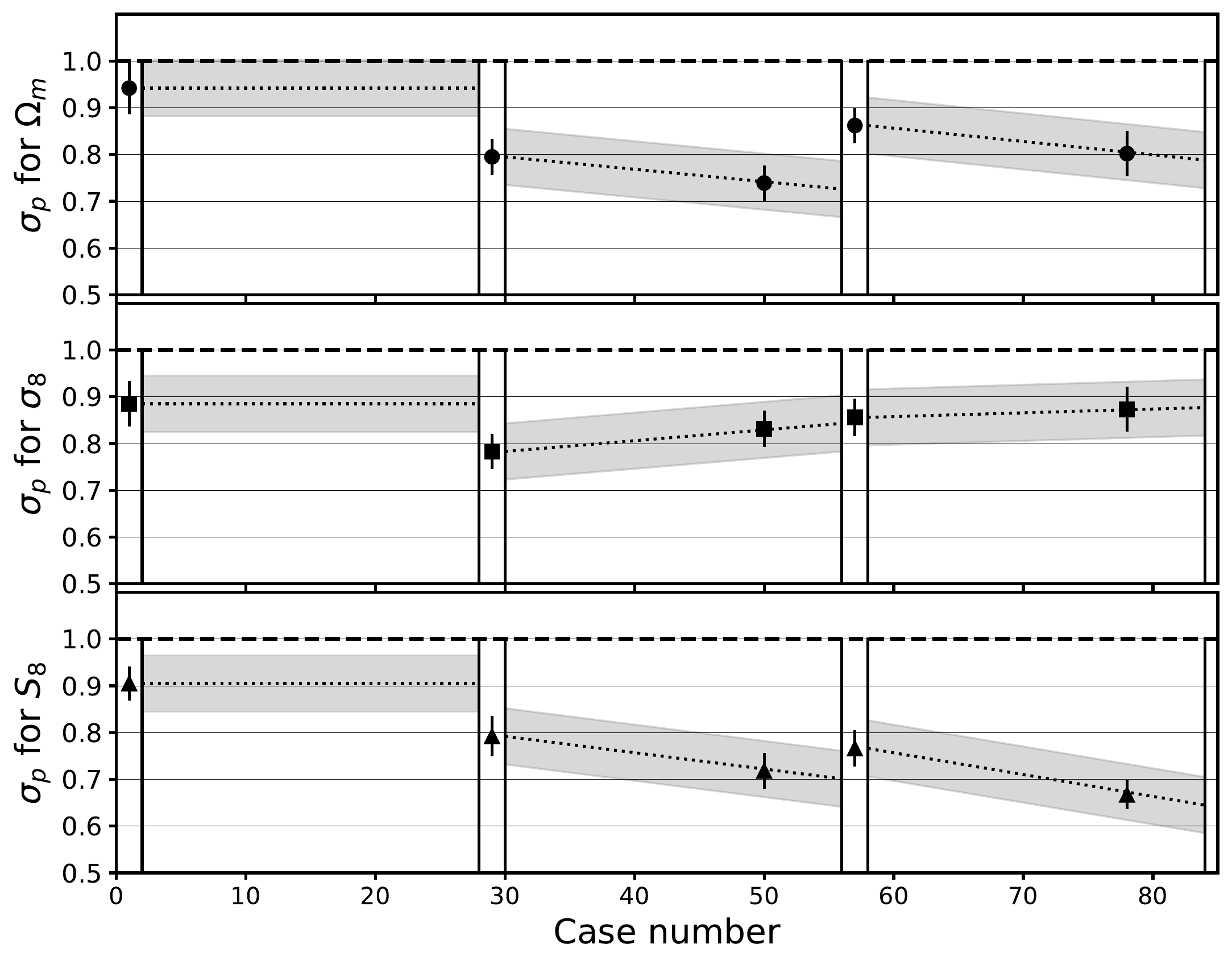}
\caption{The points with error bars show the width of the pulls $\sigma_p$ from the fits in Figure \ref{fig:pull_plot}, which are also displayed in the middle plot in Figure \ref{fig:pull_comparison} of Appendix \ref{appendix:KDE_post}.  The horizontal axis is the same as in Figure \ref{fig:peak_bias}. The dotted lines with the grey error bands show the $\sigma_p$ values used to estimate the $P_{68}$ probabilities shown in Figure \ref{fig:P68_all}.}
\label{fig:pull_sigma_bands}
\end{figure}

\begin{figure*}[htb]
\includegraphics*[width=0.7\textwidth]{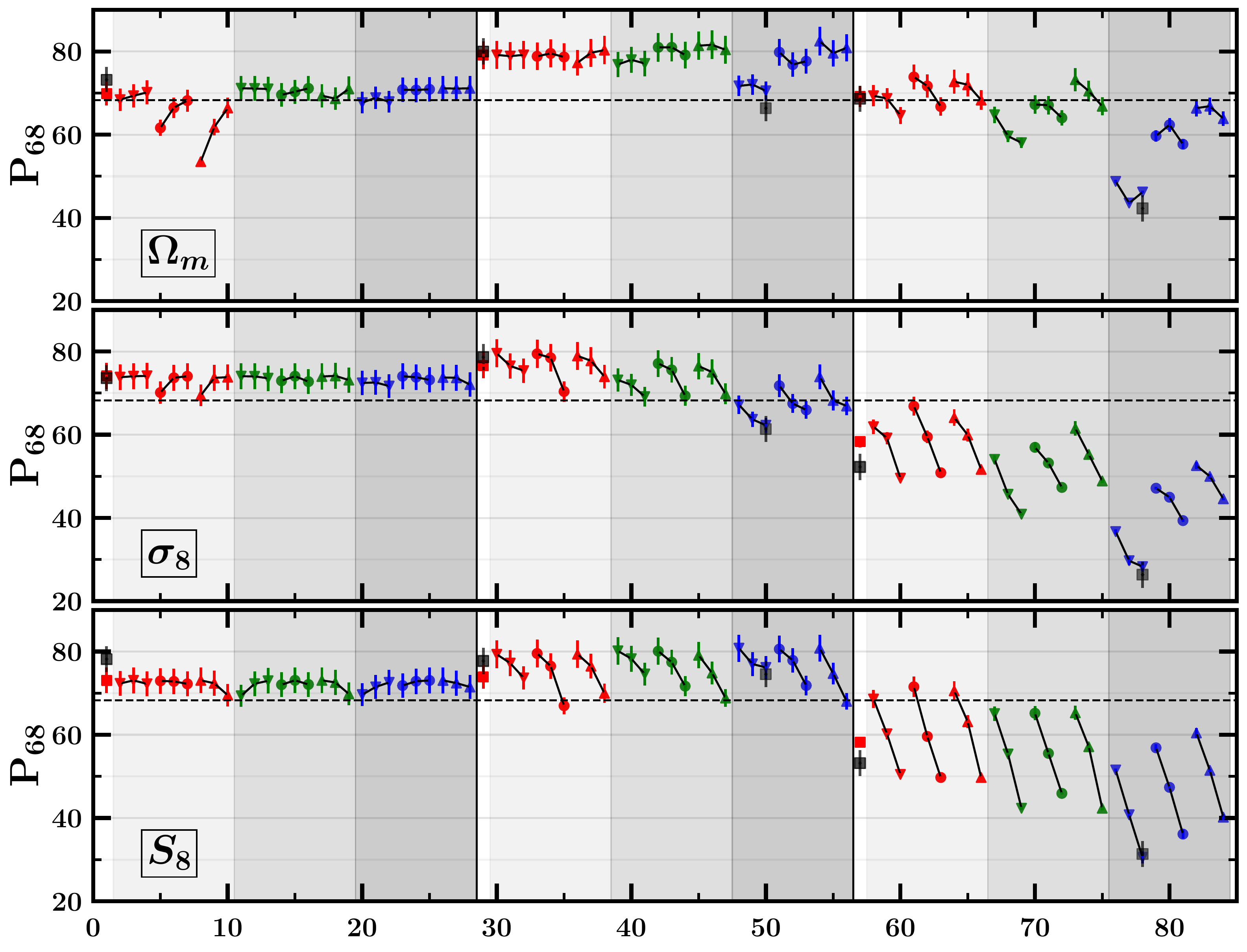}
\caption{Best estimate of the $P_{68}$ probability for the 84 combinations of parameter true values given in Tables \ref{tab:parameters} and \ref{tab:variables}.  $P_{68}$ is the probability of finding the true value of a parameter inside the 68.27\% of the area of the posterior.  See text for details of the $P_{68}$ calculation.  The horizontal order, marker and shade rules are the same as in Figure \ref{fig:peak_bias}.  From top to bottom the plots correspond to $\Omega_m$, $\sigma_8$ and $S_8$.  The five solid black squares with error bars show the values of $P_{68}$ obtained by counting, values which are also shown in Figure \ref{fig:pull_plot} and in the lower plot in Figure \ref{fig:pull_comparison} of Appendix \ref{appendix:KDE_post}.}
\label{fig:P68_all}
\end{figure*}

The results in Table \ref{tab:p68_check} clearly show that it is possible to obtain $P_{68}$ using Equation \ref{eq:p68_pull} and the pull parameters $\bar{x}_p$ and $\sigma_p$.  So we can estimate $P_{68}$ this way for all the cases in which we calculated the projection biases using synthetic data vectors but we didn't perform ensemble tests.  The comparisons shown if Figure \ref{fig:posteriors} clearly indicate that the peak of the posteriors obtained using synthetic data vectors agree very well with the peaks of the ensemble tests pull distributions.  Then to a very good approximation we can calculate the peak of the pull distribution $\bar{x}_p$ as (Posterior Peak - True value) divided by $\sigma_+$ (or $\sigma_-$) if the peak of the posterior is smaller (or greater) than the true value.  These quantities have already been plotted in Figure \ref{fig:peak_bias} for all the combinations of parameter true values given in Tables \ref{tab:parameters} and \ref{tab:variables}.  To calculate the pull distribution width $\sigma_p$ we have to make some assumptions.  Fortunately, as we will soon see, for the important case in which the Bayesian credible intervals undercover the calculation of $P_{68}$ is fairly insensitive to the value of $\sigma_p$.

The solid markers with error bars in Figure \ref{fig:pull_sigma_bands} show the values of $\sigma_p$ for the five case where we performed ensemble tests, ordered this time as in Figure \ref{fig:peak_bias}.  As in Figure \ref{fig:peak_bias} the three blocks in Figure \ref{fig:pull_sigma_bands} correspond to three different values of $\Omega_\nu h^2$.  We will assume that in each of these three blocks the behaviour of $\sigma_p$ follows an average linear behaviour similar to that of the biases in the posterior peaks observed Figure \ref{fig:peak_bias}.  We will therefore assume that $\sigma_p$ follows the dotted lines in Figure \ref{fig:pull_sigma_bands}.  For the first block we only have one ensemble test point so we assumed a horizontal line, which approximately resembles the average behaviour of the peak biases in this block.  For the error in $\sigma_p$, shown by the grey error band in Figure \ref{fig:pull_sigma_bands}, we selected the value of 0.06 which as we can see in Figure \ref{fig:pull_plot} is larger than the largest error in the values of $\sigma_p$ obtained doing ensemble tests.  

Figure \ref{fig:P68_all} shows the $P_{68}$ calculations for all the values in Figure \ref{fig:peak_bias} using Equation \ref{eq:p68_pull}.  The errors in this figure were calculated by adding $\pm 0.06$ to $\sigma_p$ in Equation \ref{eq:p68_pull}.  We can see that when the credible intervals undercover the values of $P_{68}$ are fairly independent of the value of $\sigma_p$.  This can be easily understood as follows: imagine a posterior that has a one $\sigma$ bias, then to calculate $P_{68}$ we will have to integrate between zero and two $\sigma$ and this integral will give a value very close to 0.5 independent of the width of the distribution.  The results in Figure \ref{fig:P68_all} show that when the value of $\Omega_\nu h^2$ is in the low (middle) part of the prior's range, then the 68.27\% credible intervals mostly undercover (overcover) and therefore careful coverage studies should be done before treating the credible intervals obtained in weak lensing analysis as confidence intervals.

To summarize, the studies in Section \ref{sec:ProjectionBias} clearly show that for most WL analysis, 1) there could be strong biases in the peak of the posteriors, 2) that there is a strong possibility that the errors obtained from the posteriors are inflated and 3) that there could be a strong bias in the number of times the true value falls inside the 68.27\% credible interval calculated using the posterior distributions.  The question now is how will this change with an increase in statistics, or sky coverage, in weak lensing analysis.  This is the subject of the following section.


\section{Extrapolating to larger data sets}\label{sec:LargerStatistics}

\begin{figure*}[hbt]
\includegraphics*[width=0.8\textwidth]{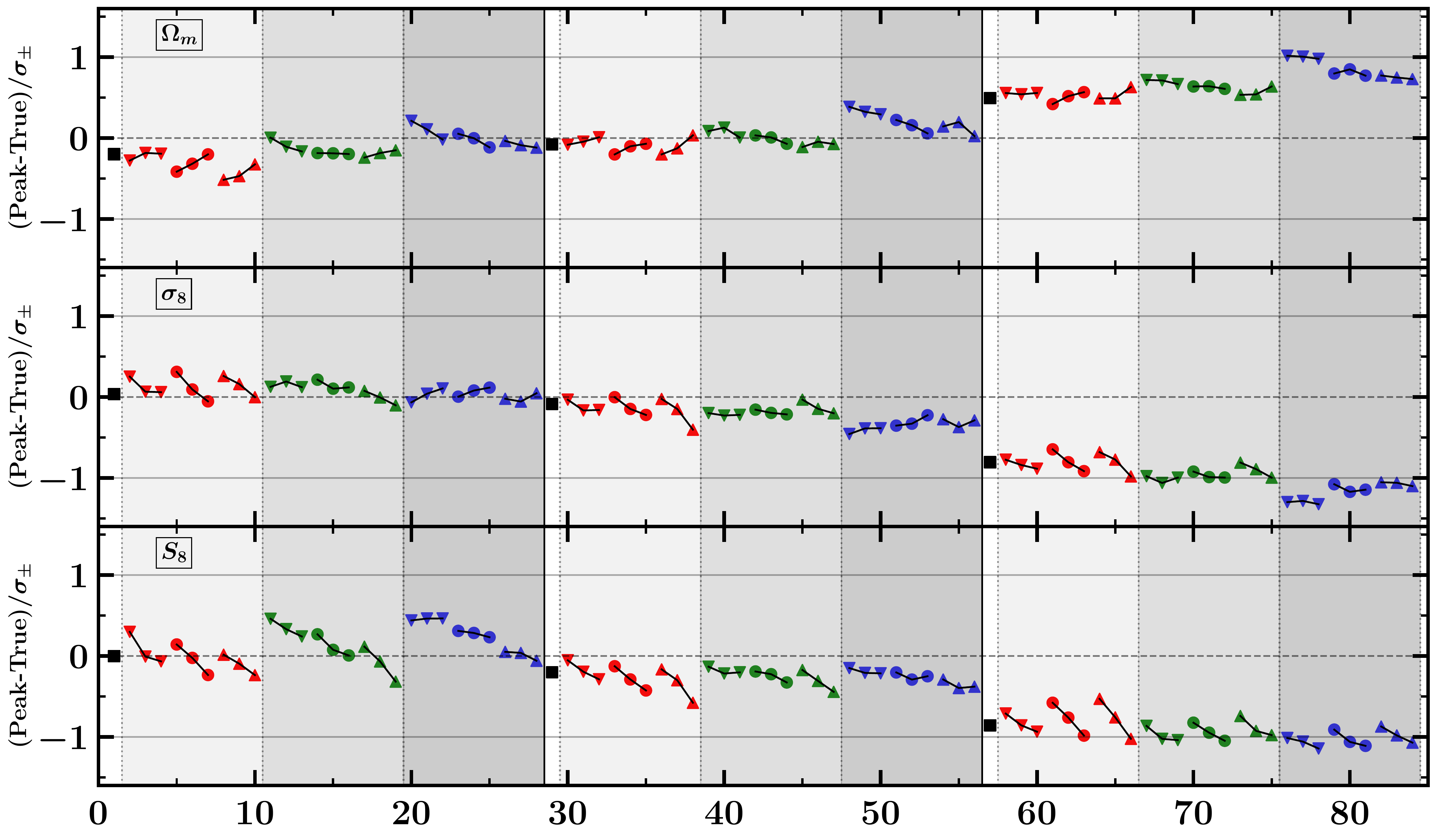}
\caption{Same set of plots as shown in Figure \ref{fig:peak_bias} but for an analysis reducing the covariance matrix of the fit to the three two-point correlation functions by a factor of three.  The factor of three was selected to simulate the recently released DES Y3 Weak Lensing analysis.}
\label{fig:peak_bias_3}
\end{figure*}
\begin{figure*}[hbt]
\includegraphics*[width=0.8\textwidth]{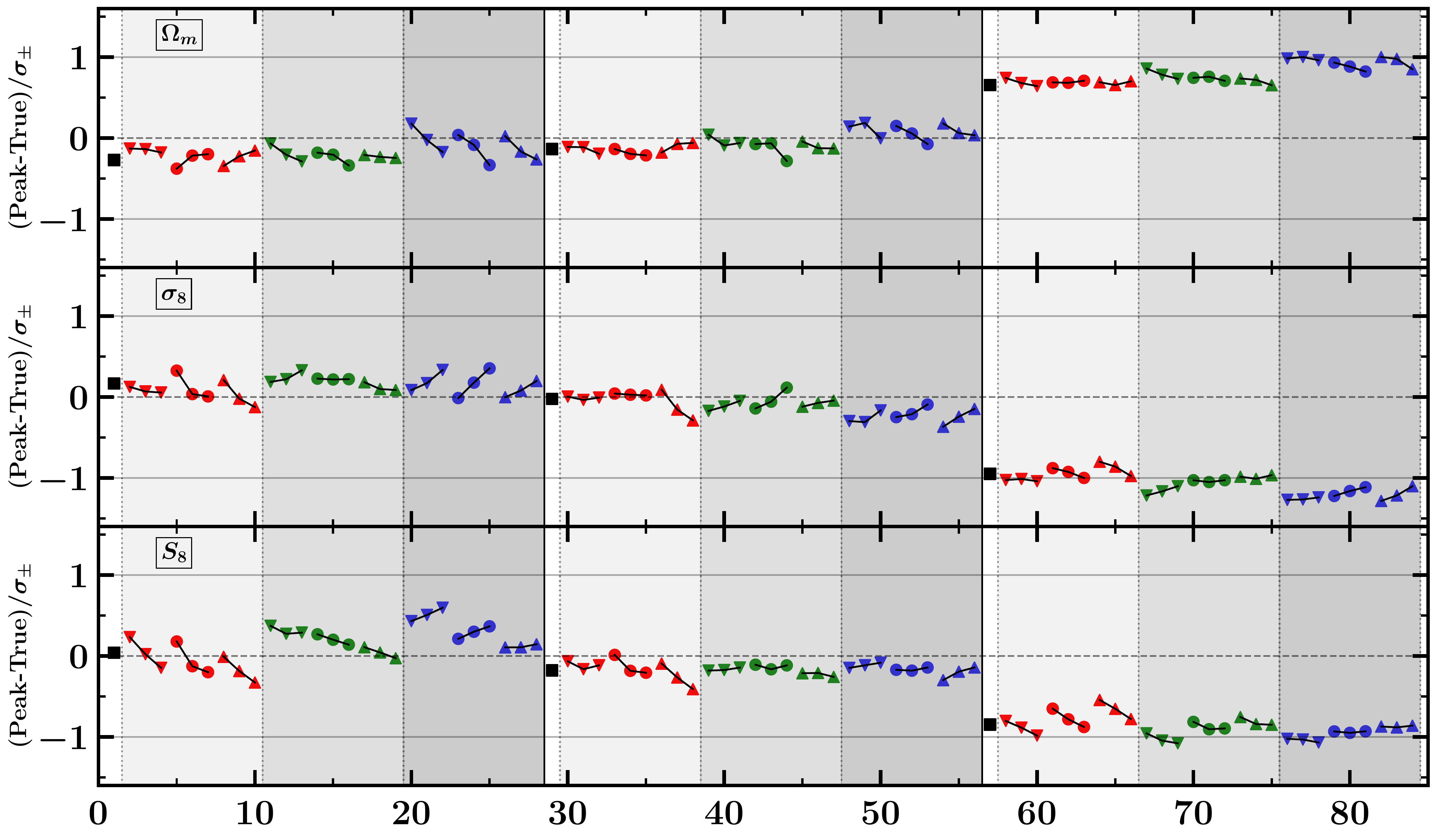}
\caption{Same set of plots as shown in Figure \ref{fig:peak_bias} but for an analysis reducing the covariance matrix of the fit to the three two-point correlation functions by a factor of nine.  The factor of nine was selected to simulate an optimistic increase in the statistics of the full analysis of the six years of DES data.}
\label{fig:peak_bias_9}
\end{figure*}

In Section \ref{sec:ProjectionBias} we demonstrated that projection biases depend on where the true values of the parameters lie inside the prior's intervals.  But we saw in Subsection \ref{subsec:example} that the projection biases also depend on how well the likelihood fits inside the ranges specified for the priors.  If the likelihood fits well inside the prior's ranges then the projection biases essentially disappear.  Then it is relevant to ask what will happen to the weak lensing analysis as the statistics, or the sky coverage and depth of the experiments increases, or the analysis methods improve.  To give a first order answer to this question we simulated a factor of three and a factor of nine increase in statistics by reducing the two-point correlation covariance matrix by a factor of three and nine respectively.  The factor of three increase in statistics intends to simulate the recent published results by DES using the data collected during the first three years of running as opposed to the first year analysis that we used in this paper.  The factor of nine is an optimistic estimation of the increase in statistics for the final DES analysis with their six years of data taking.  We have assumed that on top of the factor of six increase in statistics there will other improvements in the object reconstruction and the analysis that will produce an equivalent gain of a factor of nine increase in statistics.

Figures \ref{fig:peak_bias_3} shows the results of the projection biases in the peak of the posteriors when the two-point correlation matrix was divided by a factor of three, and Figure \ref{fig:peak_bias_9} when it was divided by a factor of nine.  In both cases the studies were done using synthetic data vectors as in Section \ref{subsec:peaks}.  An examination of Figures \ref{fig:peak_bias}, \ref{fig:peak_bias_3} and \ref{fig:peak_bias_9} show that when the true values of $n_s$, $\Omega_b$, $h$ and $\Omega_\nu h^2$ are close to the middle of the priors range (center group of 27 points) the projection biases essentially disappear with the increase in statistics.  When the true value of $\Omega_\nu h^2$ is close, or at, the minimum range (right group of 27 points) then with the increase in statistics the projection biases become more or less independent of the values of the other three parameters with biases being at or approaching the 1$\sigma$ level.  When the true value of $\Omega_\nu h^2$ is close the maximum range (left group of 27 points) then the pattern is more confusing, where the biases in $\Omega_m$ and $\sigma_8$ seem to become less dependent on  $n_s$, $\Omega_b$ and $h$ and the biases in $S_8$ seem to be more or less the same with an increase in statistics.

We do not provide an estimate of $P_{68}$ for the simulated increase in statistics because we did not perform ensemble tests in this case and therefore we don't have an estimate of the width of the pulls $\sigma_p$.  But, based on the results in Section \ref{subsec:intervals} and in Figure \ref{fig:P68_all}, it is reasonable to assume that at least for low values of the neutrino mass the credible intervals obtained in weak lensing analysis will continue to undercover.

Studies using only cosmic shear, or 1x2pt analysis, and the ones using galaxy clustering and galaxy-galaxy lensing, or 2x2pt analysis, have less constraining power than the analysis using all three two-point correlation functions, as the 3x2pt analysis used in this paper.  Less constraining power means wider likelihoods which can translate into larger projection biases \cite{Studies_1-2x2pt}.  The same is true for weak lensing analysis in the $w$CDM framework.  So even if the increase in statistics may soften the projection biases in 3x2pt analysis this paper shows that carefull attention to projection biases is warranted even with the increase in statistics expected in future weak lensing analysis.


\section{Conclusions}\label{sec:conclusions}

Given that is not possible to know the position of the true value of the parameters relative to their priors, in the current way of performing weak lensing analysis we should expect biases in the projected posterior distributions even when the analysis theory perfectly reflects Nature and there are no measurement errors.  For example, for the DES Y3 WL analysis, and only one set of true parameters, these biases can be found in Reference \cite{DES_Y3_3x2pt_validation}.  In this paper we showed that these biases arise when two conditions are satisfied: 1) we are projecting, or marginalizing, over poorly constrained parameters that are correlated with the parameters that we want to measure, and 2) the true value of these poorly constrained parameters is displaced from the center of their prior intervals.  In our case the poorly constrained parameters are $h$, $n_s$, $\Omega_b$ and $\Omega_\nu h^2$, and the biased posterior distributions correspond to $\Omega_m$, $\sigma_8$ and $S_8$.  Moreover we showed that, depending on how the poorly constrained parameters are correlated with the parameters we are interested in measuring, the projection biases can reinforce or weaken each others effects.  For example, we showed that for the DES Y1 WL analysis the maximum reinforcement occurs when the true values of $h$ and $n_s$ are close to the top of the range of their priors and the true values of $\Omega_b$ and $\Omega_\nu$ are close to the bottom of the range imposed by their priors.  
We also showed that projection biases not only affect the shift of the posterior distributions but also their widths, and that to first order we can assume that the larger the biases in the peak of the posteriors the smaller the coverage of the credible intervals, and that this will happen even as the errors become more inflated.

We could ask now the obvious question about how to correct these projection biases.  Ideally one would include run by run corrections in the analysis procedure such that ensemble tests would produce an unbiased pull distribution.  In the absence of such a procedure the best we can do is to use a well informed guess for the true values of the poorly constrained parameters and apply corrections as the ones studied in this paper.  For example if we use the PDG \cite{Zyla:2020zbs} compilation for our true values, that is $h=0.674$, $n_s=0.965$ and $\Omega_b=0.0493$ and for $\Omega_\nu h^2$ we use the minimum allowed by neutrino oscillation experiments, which corresponds to (34,48,48,1.2) per cent of their ranges respectively, then we see that we should apply the corrections for the case listed in the first row of Table \ref{tab:ensembles}.  This  case corresponds to the rightmost square symbol in Figure \ref{fig:peak_bias} and to the first row of plots in Figure \ref{fig:pull_plot}.  Then the corrections for the central values of ($\Omega_m,\sigma_8,S_8$) are ($-0.45\sigma_-,0.78\sigma_+,0.82\sigma_+$) and the multiplicative factors for correcting the errors are ($0.86,0.86,0.77$).  These corrections apply to the values in row four of Table \ref{tab:y1results}, and the corrected values are
\begin{equation}
\nonumber
\Omega_m = 0.260 ^{+0.028}_{-0.023}, \;\; \sigma_8 = 0.856 ^{+0.051}_{-0.048}, \;\; S_8 = 0.798 ^{+0.020}_{-0.018}
\end{equation}
Even though our corrections for the analysis of DES Y3 data are estimates, a comparison of Figures \ref{fig:peak_bias} and \ref{fig:peak_bias_3} for the case described above indicate that these correction probably will not change much between the Y1 and Y3 weak lensing analysis.  So to make corrections for the Y3 central values we can apply the corrections derived from the rightmost squares in Figure \ref{fig:peak_bias_3} and for the errors the corrections derived from the top row pull plots in Figure \ref{fig:pull_plot}.  Then the corrections for the central values of the posteriors will be ($-0.49\sigma_-,0.81\sigma_+,0.86\sigma_+$) and the correction factors for the errors will be the same as in the previous case ($0.86,0.86,0.77$).  Applying these corrections to the recently published DES Y3 WL results (see Table II and Figure 7 in Reference \cite{DES_Y3_3x2pt}) we get
\begin{equation}
\nonumber
\Omega_m = 0.324 ^{+0.028}_{-0.027}, \;\; \sigma_8 = 0.764 ^{+0.034}_{-0.042}, \;\; S_8 = 0.791 ^{+0.013}_{-0.013}
\end{equation}
These corrections produce smaller error bars and bring the new DES $\Lambda$CDM results to be in complete agreement with the Planck results.  We should note that the reduction in the size of the error bars coming from the pull tests is not negligible.  For example the largest reduction in $\sigma$ going from DES Y1 to Y3 is a factor of 0.74 for $S_8$.  While from the pull tests for $S_8$ we get a reduction in $\sigma$ of $0.77\pm0.04$ for $\sum m_\nu = 0.06$ eV and $0.79\pm0.04$ for $\sum m_\nu = 0.42$ eV (see the two top rightmost plots in Figure \ref{fig:pull_plot}), with the current PDG limit for the sum of the neutrino masses being $\sum m_\nu < 0.12$ eV.  So we see that a careful study of the coverage of the weak lensing analysis credible intervals could produce gains equivalent to almost doubling the amount of data available to the experiment.

As we saw, for example in Section \ref{subsec:example}, the projection biases disappear if the likelihood distribution is very well contained inside all of the allowed intervals specified by the priors.  This may lead one to believe that given the large increase in statistics of future weak lensing experiments projection biases will cease to be a problem in the future.  But the field improves as a whole, therefore the ``external" information improves too and tightening priors in poorly constrained parameters usually reduces the errors in the parameters that we are interested in measuring.  Therefore there will always be a push for tightening priors and most likely all future weak lensing experiment will have to consider paying close attention to the behaviour of their projection biases.

On the other hand, a combined analysis with different data sets able to constrain the wide posterior parameters present in WL analysis could ameliorate, or eliminate, the projection biases described in this paper.  However, if projected distributions are combined, then possibly the best way to reduce WL projection biases would be to use the best values of $h$, $n_s$, $\Omega_b$ and $\Omega_\nu h^2$ provided by the other experiments, to calculate and correct before the combination the WL results for the projection biases described in this paper.


\begin{acknowledgments}
We would like to thank Joe Zuntz and Marc Paterno for useful exchanges about CosmoSIS and Alex Drlica-Wagner and Francisco Javier Sanchez for carefully reading the manuscript and providing many useful comments, which corrected some errors and improved the quality of the paper substantially.  Prudhvi Chintalapati would like to thank the Visiting Scholar Award Program of the Universities Research Association for providing the funding to visit Fermilab and be an integral part of the studies presented in this paper, and also thank Dr. Swapan Chattopadhyay of Northern Illinois University for his financial support.  This manuscript has been authored by Fermi Research Alliance, LLC under Contract No. DE-AC02-07CH11359 with the U.S. Department of Energy, Office of High Energy Physics.
\end{acknowledgments}


\appendix

\section{Reducing the number of $\mathbf{N_{ell}}$ points in CosmoSIS}\label{appendix:n_ell}

The default DES Y1 WL analysis version stored in CosmoSIS calculates the inner integrals in Equations \ref{eq:gal-clus} to \ref{eq:shear-shear} in the range $0.1 \le l \le 5\times 10^5$ using $N_{ell}=400$ logarithmically spaced points.  In order to reduce the amount of CPU time needed for the studies in this paper to a manageable amount (about six million CPU hours) the number of integration points was reduced from  $N_{ell}=400$ to $N_{ell}=200$.  In this Appendix we show that the effect of this change is very small.

\begin{figure}[hbt]
\includegraphics[width=0.48\textwidth]{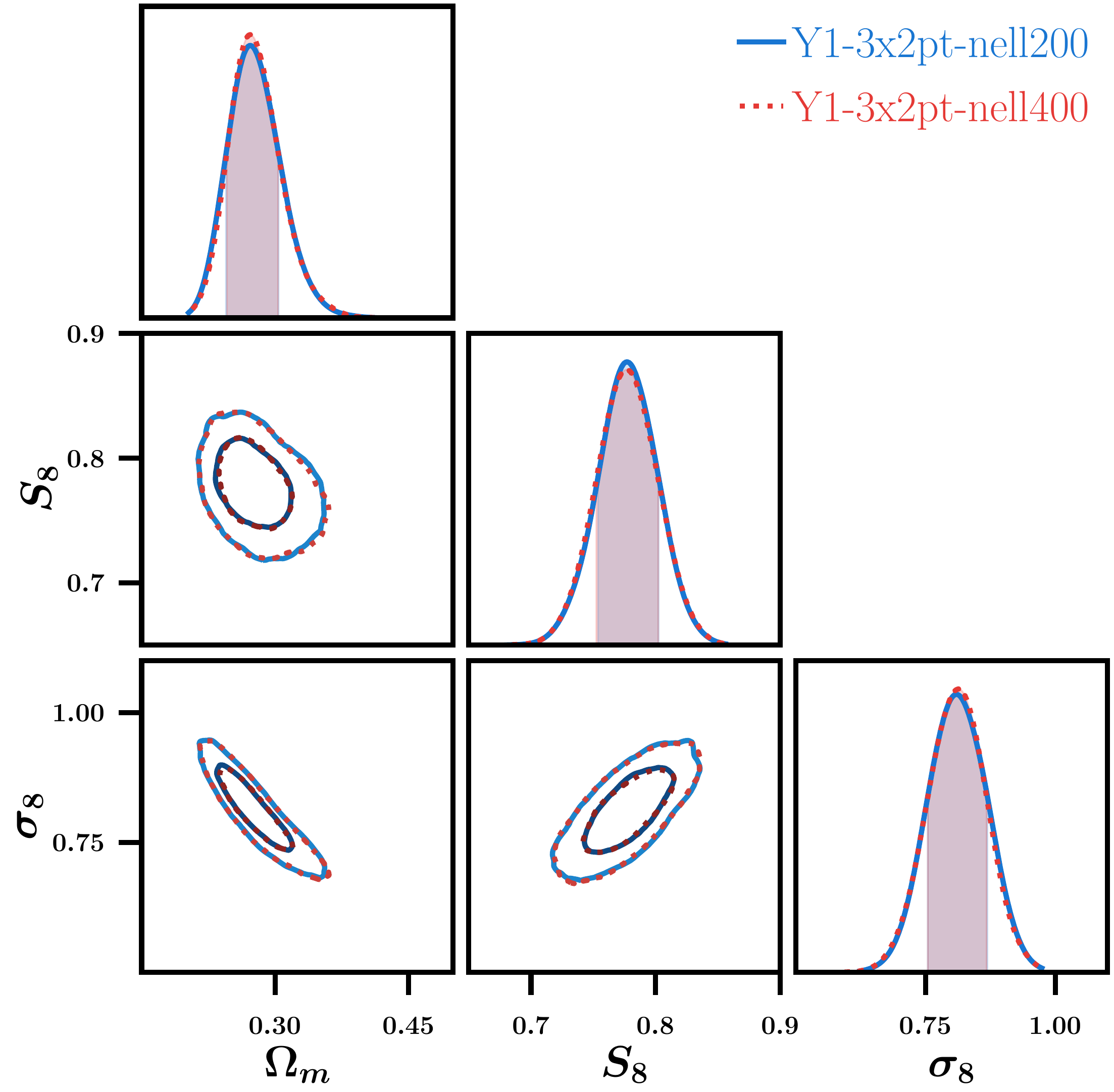}
\caption{One and two dimensional posterior plots for the analysis of the DES Y1 weak lensing data.  The solid blue (dotted red) contour corresponds to $N_{ell}$ = 200 (400).  The two cases are essentially indistinguishable.}
\label{fig:appendix_nell_comp}
\end{figure}

Figure \ref{fig:appendix_nell_comp} shows the one and two dimensional posterior distributions for the analysis of the DES Y1 weak lensing data.  The solid blue (dotted red) contour corresponds to the analysis of the data using $N_{ell}=200 \; (400)$ points in the calculation of the inner integrals in Equations \ref{eq:gal-clus} to \ref{eq:shear-shear}.  We can see that the two cases are essentially indistinguishable both in the one and two dimensional distributions.  For a detail comparison, Table \ref{tab:appendix_nell} shows the peak values of the posteriors and their errors for $\Omega_m$, $S_8$ and $\sigma_8$.  Again we see that the differences between the two analysis are very small.

\begin{table}[bht]
\renewcommand{\arraystretch}{1.5}
\caption{\label{tab:appendix_nell} Results of the analysis of the DES Y1 weak lensing data with two different values of $N_{ell}$.}
\begin{ruledtabular}
\begin{tabular}{cccc}
$N_{ell}$ &$\Omega_{m}$ & $S_{8}$& $\sigma_{8}$ \\
\colrule
200 & $0.272 ^{+0.032}_{-0.027}$ & $0.777 ^{+0.026}_{-0.023}$ & $0.810 ^{+0.059}_{-0.056}$ \\
400 & $0.272 ^{+0.032}_{-0.026}$ & $0.777 ^{+0.026}_{-0.024}$ & $0.812 ^{+0.055}_{-0.058}$ \\
\end{tabular}
\end{ruledtabular}
\end{table}

\begin{figure}[hbt]
\includegraphics[width=0.48\textwidth]{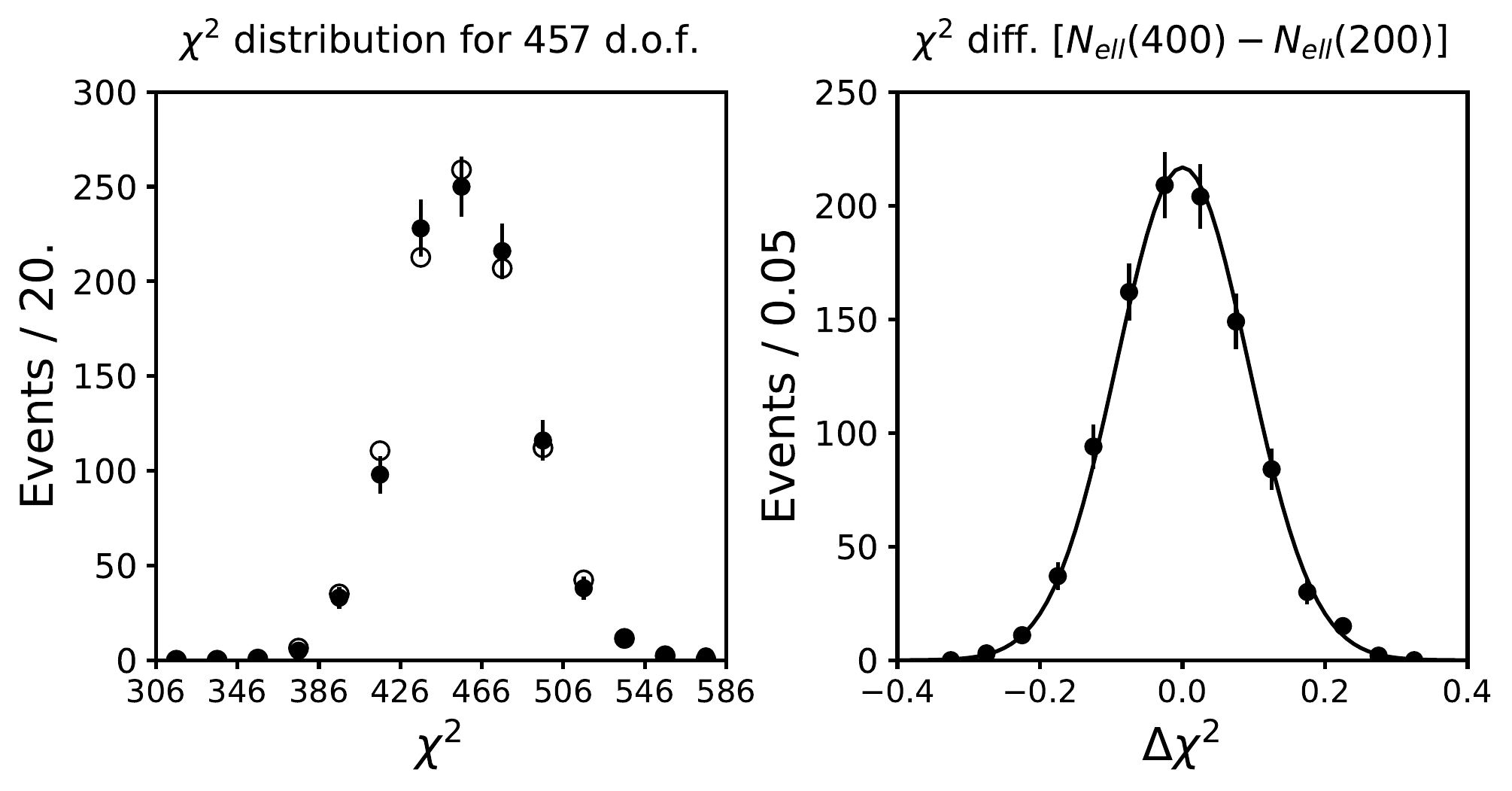}
\caption{The solid points with errors bars show the $\chi^2$ distribution for $N_{ell}=200$ (left) and the $\chi^2$ difference between the $N_{ell}=400$ and the $N_{ell}=200$ runs (right).  The open points correspond to the $\chi^2$ theoretical prediction and the solid line is a Gaussian fit giving $\sigma=0.092$.  The $\chi^2$ values were calculated using the 457 points comprising the three two point correlation functions for each of the one thousand runs using input ``data" that was generated fluctuating the theory prediction with the covariance matrix.}
\label{fig:appendix_nell_chi2_diff}
\end{figure}

To make sure that the small differences observed in the analysis of the data are not a fluctuation we also compared the values of $\chi^2$ calculated using the 457 points corresponding to the three two point correlation functions that are used in the analysis.  We produced one thousand runs where the input ``data" (the 3x2pt correlation functions) was generated fluctuating the theoretical calculation of the two point correlation function using the covariance matrix.  Then $\chi^2$ was calculated using these 457 points and the theoretical predictions using $N_{ell}=400$ and 200.  The left plot in Figure \ref{fig:appendix_nell_chi2_diff} shows the $\chi^2$ distribution for $N_{ell}=200$ and the right plot shows a run-by-run difference of $\chi^2$ for $N_{ell}=400$ and 200.  We can again see that the $\chi^2$ differences are very small in comparison to the width of the $\chi^2$ distribution, confirming that using $N_{ell}=200$ points instead of 400 points to calculate the inner integrals in Equations \ref{eq:gal-clus} to \ref{eq:shear-shear} have little effect in the overall analysis.

\section{Minimum Credible Intervals}\label{appendix:confidence_int}

Given a normalized posterior $P(x)$ we want to calculate two limits $A$ and $B$ such that
\begin{equation}
\int_A^B P(x) \, dx = a_{68}
\label{eqA:area_68}
\end{equation}

\vspace{0.25cm} \noindent
where $a_{68} = \mbox{erf}(1/\sqrt{2}) \approx 0.6827$.  Equation \ref{eqA:area_68} is not enough to determine the values of $A$ and $B$, another condition is needed.  Here we will impose the condition that the interval $B-A$ be minimal.  Then the problem we want to solve is the minimization of the interval $B-A$ subject to condition \ref{eqA:area_68}.  Using a Lagrange multiplier we need to minimize
\begin{equation}
Q = B - A + \lambda \left( \int_A^B P(x) \, dx - a_{68} \right)
\label{eqA:lagrange}
\end{equation}
Given the minimization conditions
\begin{equation}
\frac{\partial Q}{\partial B} = \frac{\partial Q}{\partial A} =\frac{\partial Q}{\partial \lambda} = 0
\end{equation}
we obtain
\begin{equation}
1+\lambda \, P(B) = 0 \;,\; \mbox{ and } \;\; -1-\lambda \, P(A) = 0
\end{equation}
plus Equation \ref{eqA:area_68}, which leads to the relation
\begin{equation}
P(A) = P(B)
\label{eq:CI_equality}
\end{equation}
If $P(x)$ has only one peak then the previous condition also ensures that the peak is contained in the credible interval.  This happens because $P(x)$ steadily falls on both sides of the peak.  If $P(x)$ has more that one peak then both peaks could be inside of the credible interval, or condition \ref{eq:CI_equality} could give rise to two credible intervals and one would select the smaller of the two.

\section{Calculating $\mathbf{P_{68}}$ From Pull Parameters}\label{appendix:P68_pull}

\begin{figure}[hbt]
\vspace{0.8cm}
\includegraphics[width=0.48\textwidth]{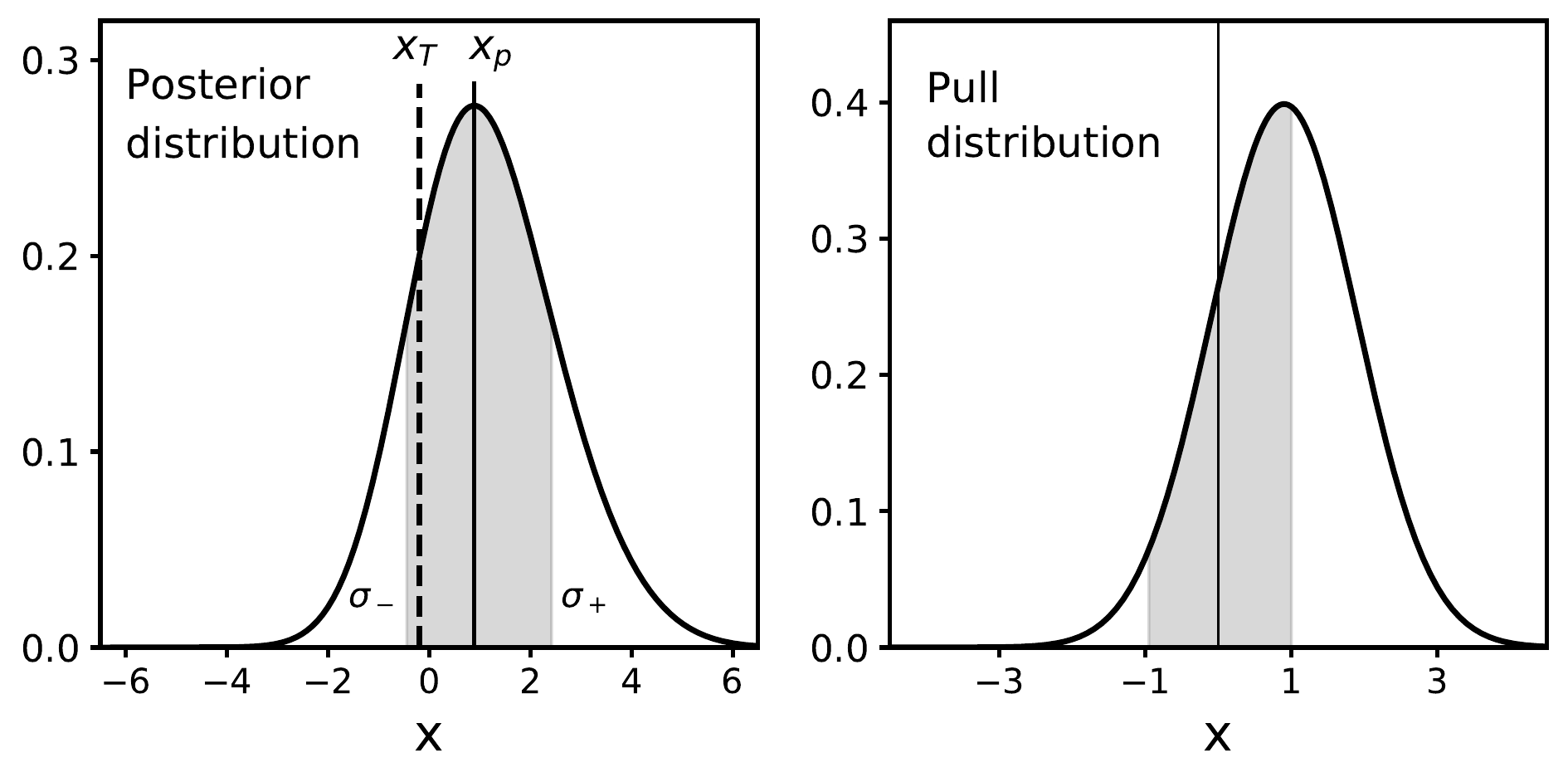}
\vspace{-0.3cm}
\caption{Conceptual posterior distribution for a single experiment (left) and pull distribution (right).  For a large number of experiments the 68.27\% area under the posteriors translates into the [-1,1] area under the pull distribution.}
\label{fig:P68_pull}
\end{figure}

The left plot in Figure \ref{fig:P68_pull} shows a conceptual posterior distribution for one experiment together with the peak of the distribution $x_p$ and the true value $x_T$.  
The true value lies inside the 68.27\% credible interval if the condition $x_p - \sigma_- \le x_T \le x_p + \sigma_+$ is satisfied, which translates into the inequalities
\begin{equation}
-\sigma_+ \le x_p - x_T \le \sigma_-
\label{eq:p68_pull_1}
\end{equation}
For a very large number of experiment we can form the pull defined as in Equation \ref{eq:pull_definition}.  When $x_T$ is close to $\sigma_-$ we have to divide $x_p - x_T$ by $\sigma_-$ and the right inequality in Equation \ref{eq:p68_pull_1} becomes $\mbox{pull} \le 1$, and when $x_T$ is close to $\sigma_+$ the left inequality in the same equation becomes $\mbox{pull} \ge -1$.   Therefore for a large number of experiments the condition for finding the true value inside the 68.27\% area of the posterior translates into $-1 \le \mbox{pull} \le 1$, as illustrated in the right plot of Figure \ref{fig:P68_pull}.  Then for a Gaussian pull the probability of finding the true value inside the 68.27\% area of the posterior is given by
\begin{align}
P_{68} &= \frac{1}{\sqrt{2\pi}\sigma_p} \int_{-1}^1 dx \, e^{-\frac{1}{2}\left( \frac{x-\bar{x}_p}{\sigma_p} \right)^2} \\
&= \frac{1}{2} \left\{ \mbox{erf}\left(\frac{1-\bar{x}_p}{\sqrt{2}\,\sigma_p}\right) + \mbox{erf}\left(\frac{1+\bar{x}_p}{\sqrt{2}\,\sigma_p}\right)  \right\}
\end{align}
where $\bar{x}_p$ and $\sigma_p$ are the mean and rms of the pull distribution and we have used the property $\mbox{erf}(-x) = -\mbox{erf}(x)$.  The error in $P_{68}$ is calculated in the standard way
\begin{equation}
\sigma_{P_{68}} = \sqrt{ \left( \frac{\partial P_{68}}{\partial \bar{x}_p} \, \Delta \bar{x}_p  \right)^2 + \left( \frac{\partial P_{68}}{\partial \sigma_p} \, \Delta \sigma_p  \right)^2 }
\end{equation}

\section{KDE and Posterior Widths}\label{appendix:KDE_post}

The results in Section \ref{subsec:intervals}, and the simple example in Section \ref{subsec:example}, clearly show that the posterior credible intervals in weak lensing analysis can be substantially inflated.  The question is, are there other sources that might contribute to the inflation of errors other than projection biases?  Another source of error inflation is the Gaussian Kernel Density Estimator (KDE) algorithm \cite{KDE_algo} used by programs like Chainconsumer \cite{Chainconsumer_ref}.  This program is widely used in weak lensing analysis to process the output of samplers like Multinest or Markov chain Monte Carlo programs.  In the Gaussian KDE algorithm each point is replaced by a finite width Gaussian distribution.  In Chainconsumer, the width of the Gaussian is controlled by a parameter called KDE.
\begin{figure}[hbt]
\includegraphics[width=0.48\textwidth]{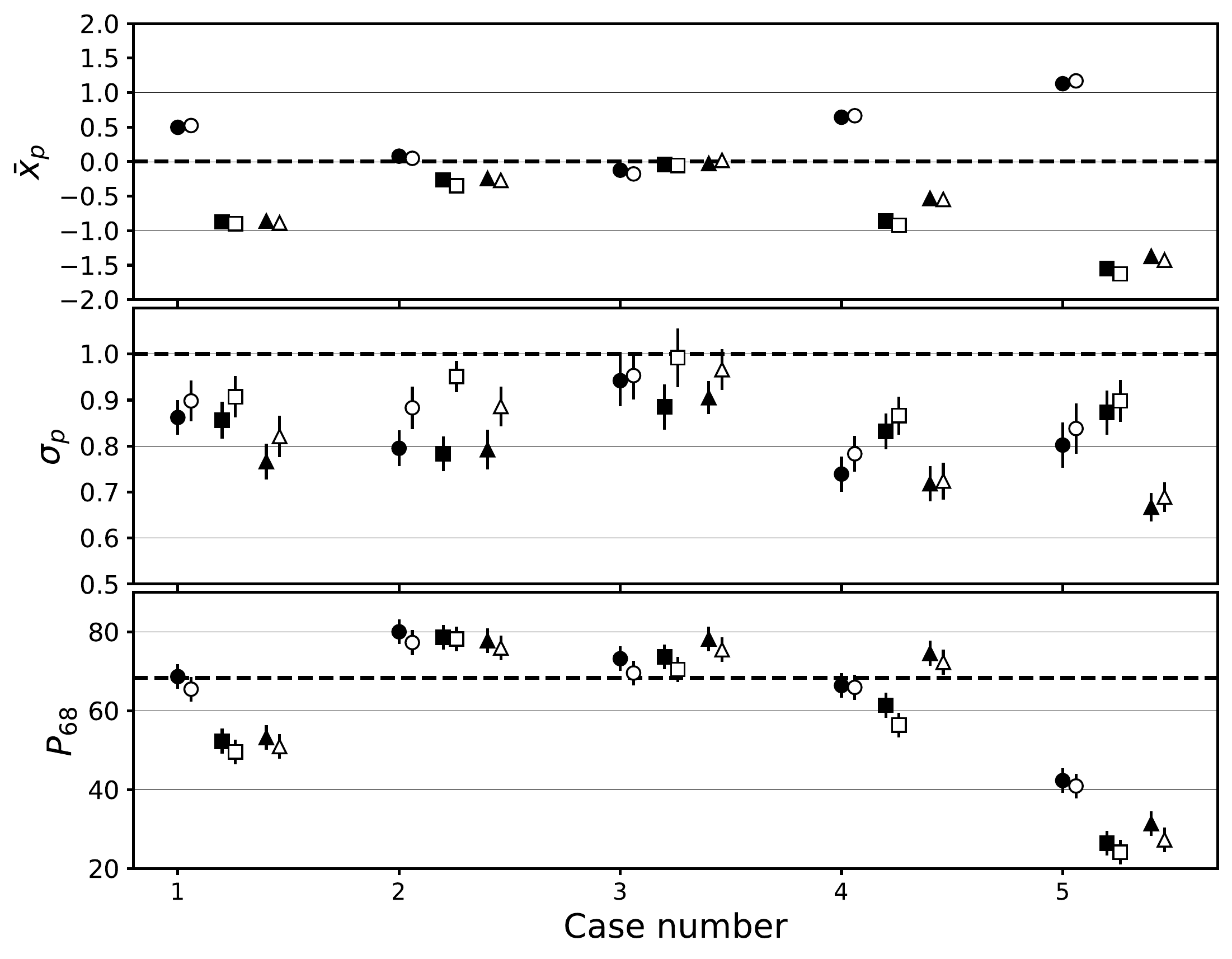}
\caption{Plots of the pull mean $\bar{x}_p$ and rms $\sigma_p$ and the 68.27\% probability $P_{68}$.  The circles, squares and triangles correspond to $\Omega_m$, $\sigma_8$ and $S_8$ respectively.  The solid markers correspond to KDE=1.5 and the open ones to KDE=1.0.  The case numbers are the same as in Table \ref{tab:ensembles} or from left to right correspond to the top to bottom plots in Figure \ref{fig:pull_plot}.  For each case the markers have been displaced for clarity.  The dashed lines give the expected values in the absence of biases.}
\label{fig:pull_comparison}
\end{figure}

Figure \ref{fig:pull_comparison} shows the comparison between two analyses of all five ensemble tests using two different values of KDE in Chainconsumer.  The solid markers show the pull parameters $\bar{x}_p$ and $\sigma_p$ and the probability $P_{68}$ for KDE=1.5.  All these values come from the plots in Figure \ref{fig:pull_plot}.  The circles, squares and triangles correspond to $\Omega_m$, $\sigma_8$ and $S_8$ and from left to right the Case number corresponds to the top to bottom plots in Figure \ref{fig:pull_plot}.  The open markers correspond to a reanalysis of all five ensemble tests using KDE=1.0.  As expected a lower value of KDE produces narrower posterior and therefore smaller credible intervals, which translates to larger values of $\sigma_p$ and smaller values of $P_{68}$.  For Cases 1, 4 and 5 the changes in $\sigma_p$ are small and almost within error bars.  The changes in $\sigma_p$ are larger for Case 2, but this could be a fluctuation in the Gaussian fits to the pull distribution (the fits are unstable in this case) because the changes in the more stable $P_{68}$ are the same for Case 2 as they are for all the other cases.  So we see an improvement in the credible intervals going from KDE=1.5 to the marginally low value of KDE=1.0, but it is not enough to account for the inflated values of the credible intervals we observe in most of the ensemble tests.

In principle it is possible to go to lower values of KDE but this requires increasing the number of points used to calculate posterior distributions.  In samplers like Multinest this is controlled by the number of \emph{live-points}.  The number used in this paper (500) is the same as that used, for example, in the recently published DES Weak Lensing analysis using the first three years of data.  This number of live-points typically produces about 5,000 points with weights large enough to contribute to the posterior distributions, but this number of points is not enough to produce smooth 95\% credible range edges for two dimensional posteriors using KDE values of one or lower.  Increasing the live-points by a factor of $n$ (e.g. $n=$ 2,4,8,...) increases the number of points in the posteriors by a factor of $n$ but also increases the CPU time by the same $n$ factor.  The studies presented in this paper took about five months of running, almost exclusively, on 14 computer nodes with a total of 1728 CPUs, so it is not really an option to increase the CPU time by factors of two.


\bibliography{PRD_PGM}

\providecommand{\noopsort}[1]{}\providecommand{\singleletter}[1]{#1}%
\begin{thebibliography}{43}%
\makeatletter
\providecommand \@ifxundefined [1]{%
 \@ifx{#1\undefined}
}%
\providecommand \@ifnum [1]{%
 \ifnum #1\expandafter \@firstoftwo
 \else \expandafter \@secondoftwo
 \fi
}%
\providecommand \@ifx [1]{%
 \ifx #1\expandafter \@firstoftwo
 \else \expandafter \@secondoftwo
 \fi
}%
\providecommand \natexlab [1]{#1}%
\providecommand \enquote  [1]{``#1''}%
\providecommand \bibnamefont  [1]{#1}%
\providecommand \bibfnamefont [1]{#1}%
\providecommand \citenamefont [1]{#1}%
\providecommand \href@noop [0]{\@secondoftwo}%
\providecommand \href [0]{\begingroup \@sanitize@url \@href}%
\providecommand \@href[1]{\@@startlink{#1}\@@href}%
\providecommand \@@href[1]{\endgroup#1\@@endlink}%
\providecommand \@sanitize@url [0]{\catcode `\\12\catcode `\$12\catcode
  `\&12\catcode `\#12\catcode `\^12\catcode `\_12\catcode `\%12\relax}%
\providecommand \@@startlink[1]{}%
\providecommand \@@endlink[0]{}%
\providecommand \url  [0]{\begingroup\@sanitize@url \@url }%
\providecommand \@url [1]{\endgroup\@href {#1}{\urlprefix }}%
\providecommand \urlprefix  [0]{URL }%
\providecommand \Eprint [0]{\href }%
\providecommand \doibase [0]{https://doi.org/}%
\providecommand \selectlanguage [0]{\@gobble}%
\providecommand \bibinfo  [0]{\@secondoftwo}%
\providecommand \bibfield  [0]{\@secondoftwo}%
\providecommand \translation [1]{[#1]}%
\providecommand \BibitemOpen [0]{}%
\providecommand \bibitemStop [0]{}%
\providecommand \bibitemNoStop [0]{.\EOS\space}%
\providecommand \EOS [0]{\spacefactor3000\relax}%
\providecommand \BibitemShut  [1]{\csname bibitem#1\endcsname}%
\let\auto@bib@innerbib\@empty
\bibitem [{\citenamefont {Fisher}\ \emph {et~al.}(2000)\citenamefont {Fisher}
  \emph {et~al.}}]{Sloan_WL}%
  \BibitemOpen
  \bibfield  {author} {\bibinfo {author} {\bibfnamefont {P.}~\bibnamefont
  {Fisher}} \emph {et~al.} (\bibinfo {collaboration} {SDSS}),\ }\href@noop {}
  {\bibfield  {journal} {\bibinfo  {journal} {The Astronomical Journal}\
  }\textbf {\bibinfo {volume} {120}},\ \bibinfo {pages} {1198} (\bibinfo {year}
  {2000})}\BibitemShut {NoStop}%
\bibitem [{\citenamefont {Wittman}\ \emph {et~al.}(2000)\citenamefont
  {Wittman}, \citenamefont {Tyson}, \citenamefont {Kirkman}, \citenamefont
  {Dell’Antonio},\ and\ \citenamefont {Bernstein}}]{Nature_2000_WL}%
  \BibitemOpen
  \bibfield  {author} {\bibinfo {author} {\bibfnamefont {D.~M.}\ \bibnamefont
  {Wittman}}, \bibinfo {author} {\bibfnamefont {J.~A.}\ \bibnamefont {Tyson}},
  \bibinfo {author} {\bibfnamefont {D.}~\bibnamefont {Kirkman}}, \bibinfo
  {author} {\bibfnamefont {I.}~\bibnamefont {Dell’Antonio}},\ and\ \bibinfo
  {author} {\bibfnamefont {G.}~\bibnamefont {Bernstein}},\ }\href@noop {}
  {\bibfield  {journal} {\bibinfo  {journal} {Nature}\ }\textbf {\bibinfo
  {volume} {405}},\ \bibinfo {pages} {143} (\bibinfo {year}
  {2000})}\BibitemShut {NoStop}%
\bibitem [{\citenamefont {Bacon}\ \emph {et~al.}(2000)\citenamefont {Bacon},
  \citenamefont {Refregier},\ and\ \citenamefont {Ellis}}]{MNRAS_2000_WL}%
  \BibitemOpen
  \bibfield  {author} {\bibinfo {author} {\bibfnamefont {D.~J.}\ \bibnamefont
  {Bacon}}, \bibinfo {author} {\bibfnamefont {A.~R.}\ \bibnamefont
  {Refregier}},\ and\ \bibinfo {author} {\bibfnamefont {R.~S.}\ \bibnamefont
  {Ellis}},\ }\href@noop {} {\bibfield  {journal} {\bibinfo  {journal} {MNRAS}\
  }\textbf {\bibinfo {volume} {318}},\ \bibinfo {pages} {625} (\bibinfo {year}
  {2000})}\BibitemShut {NoStop}%
\bibitem [{\citenamefont {van Waerbeke}\ \emph {et~al.}(2000)\citenamefont {van
  Waerbeke} \emph {et~al.}}]{A&A_2000_WL}%
  \BibitemOpen
  \bibfield  {author} {\bibinfo {author} {\bibfnamefont {L.}~\bibnamefont {van
  Waerbeke}} \emph {et~al.} (\bibinfo {collaboration} {CFHT}),\ }\href@noop {}
  {\bibfield  {journal} {\bibinfo  {journal} {A\&A}\ }\textbf {\bibinfo
  {volume} {358}},\ \bibinfo {pages} {30} (\bibinfo {year} {2000})}\BibitemShut
  {NoStop}%
\bibitem [{\citenamefont {Amon}\ \emph {et~al.}(2021)\citenamefont {Amon} \emph
  {et~al.}}]{DES_Y3_1x2pt-a}%
  \BibitemOpen
  \bibfield  {author} {\bibinfo {author} {\bibfnamefont {A.}~\bibnamefont
  {Amon}} \emph {et~al.} (\bibinfo {collaboration} {DES}),\ }\href@noop {}
  {\bibfield  {journal} {\bibinfo  {journal} {arXiv:2105.13543}\ } (\bibinfo
  {year} {2021})}\BibitemShut {NoStop}%
\bibitem [{\citenamefont {Secco}\ \emph {et~al.}(2021)\citenamefont {Secco}
  \emph {et~al.}}]{DES_Y3_1x2pt-b}%
  \BibitemOpen
  \bibfield  {author} {\bibinfo {author} {\bibfnamefont {L.~F.}\ \bibnamefont
  {Secco}} \emph {et~al.} (\bibinfo {collaboration} {DES}),\ }\href@noop {}
  {\bibfield  {journal} {\bibinfo  {journal} {arXiv:2105.13544}\ } (\bibinfo
  {year} {2021})}\BibitemShut {NoStop}%
\bibitem [{\citenamefont {Pandey}\ \emph {et~al.}(2021)\citenamefont {Pandey}
  \emph {et~al.}}]{DES_Y3_2x2pt-a}%
  \BibitemOpen
  \bibfield  {author} {\bibinfo {author} {\bibfnamefont {S.}~\bibnamefont
  {Pandey}} \emph {et~al.} (\bibinfo {collaboration} {DES}),\ }\href@noop {}
  {\bibfield  {journal} {\bibinfo  {journal} {arXiv:2105.13545}\ } (\bibinfo
  {year} {2021})}\BibitemShut {NoStop}%
\bibitem [{\citenamefont {Porredon}\ \emph {et~al.}(2021)\citenamefont
  {Porredon} \emph {et~al.}}]{DES_Y3_2x2pt-b}%
  \BibitemOpen
  \bibfield  {author} {\bibinfo {author} {\bibfnamefont {A.}~\bibnamefont
  {Porredon}} \emph {et~al.} (\bibinfo {collaboration} {DES}),\ }\href@noop {}
  {\bibfield  {journal} {\bibinfo  {journal} {arXiv:2105.13546}\ } (\bibinfo
  {year} {2021})}\BibitemShut {NoStop}%
\bibitem [{\citenamefont {Abbott}\ \emph {et~al.}(2021)\citenamefont {Abbott}
  \emph {et~al.}}]{DES_Y3_3x2pt}%
  \BibitemOpen
  \bibfield  {author} {\bibinfo {author} {\bibfnamefont {T.~M.~C.}\
  \bibnamefont {Abbott}} \emph {et~al.} (\bibinfo {collaboration} {DES}),\
  }\href@noop {} {\bibfield  {journal} {\bibinfo  {journal} {arXiv:2105.13549}\
  } (\bibinfo {year} {2021})}\BibitemShut {NoStop}%
\bibitem [{\citenamefont {Hamana}\ \emph {et~al.}(2020)\citenamefont {Hamana}
  \emph {et~al.}}]{HSC_2020}%
  \BibitemOpen
  \bibfield  {author} {\bibinfo {author} {\bibfnamefont {T.}~\bibnamefont
  {Hamana}} \emph {et~al.} (\bibinfo {collaboration} {HSC}),\ }\href@noop {}
  {\bibfield  {journal} {\bibinfo  {journal} {PASJ}\ }\textbf {\bibinfo
  {volume} {72}},\ \bibinfo {pages} {16} (\bibinfo {year} {2020})}\BibitemShut
  {NoStop}%
\bibitem [{\citenamefont {Asgari}\ \emph {et~al.}(2021)\citenamefont {Asgari}
  \emph {et~al.}}]{KiDS-1000_2021_1x2pt}%
  \BibitemOpen
  \bibfield  {author} {\bibinfo {author} {\bibfnamefont {M.}~\bibnamefont
  {Asgari}} \emph {et~al.} (\bibinfo {collaboration} {KiDS}),\ }\href@noop {}
  {\bibfield  {journal} {\bibinfo  {journal} {A\&A}\ }\textbf {\bibinfo
  {volume} {645}},\ \bibinfo {pages} {A104} (\bibinfo {year}
  {2021})}\BibitemShut {NoStop}%
\bibitem [{\citenamefont {Heymans}\ \emph {et~al.}(2021)\citenamefont {Heymans}
  \emph {et~al.}}]{KiDS-1000_2021_3x2pt}%
  \BibitemOpen
  \bibfield  {author} {\bibinfo {author} {\bibfnamefont {C.}~\bibnamefont
  {Heymans}} \emph {et~al.} (\bibinfo {collaboration} {KiDS}),\ }\href@noop {}
  {\bibfield  {journal} {\bibinfo  {journal} {A\&A}\ }\textbf {\bibinfo
  {volume} {646}},\ \bibinfo {pages} {A140} (\bibinfo {year}
  {2021})}\BibitemShut {NoStop}%
\bibitem [{\citenamefont {Tröster}\ \emph {et~al.}(2021)\citenamefont
  {Tröster} \emph {et~al.}}]{KiDS-1000_2021_beyondLCDM}%
  \BibitemOpen
  \bibfield  {author} {\bibinfo {author} {\bibfnamefont {T.}~\bibnamefont
  {Tröster}} \emph {et~al.} (\bibinfo {collaboration} {KiDS}),\ }\href@noop {}
  {\bibfield  {journal} {\bibinfo  {journal} {A\&A}\ }\textbf {\bibinfo
  {volume} {649}},\ \bibinfo {pages} {A88} (\bibinfo {year}
  {2021})}\BibitemShut {NoStop}%
\bibitem [{\citenamefont {Li}\ \emph {et~al.}(2021)\citenamefont {Li} \emph
  {et~al.}}]{HSC_2021_3YearCatalog}%
  \BibitemOpen
  \bibfield  {author} {\bibinfo {author} {\bibfnamefont {X.}~\bibnamefont {Li}}
  \emph {et~al.} (\bibinfo {collaboration} {HSC}),\ }\href@noop {} {\bibfield
  {journal} {\bibinfo  {journal} {arXiv:2107.00136v1}\ } (\bibinfo {year}
  {2021})}\BibitemShut {NoStop}%
\bibitem [{Note1()}]{Note1}%
  \BibitemOpen
  \bibinfo {note} {Euclid: https://www.euclid-ec.org}\BibitemShut {NoStop}%
\bibitem [{Note2()}]{Note2}%
  \BibitemOpen
  \bibinfo {note} {LSST: https://www.lsst.org}\BibitemShut {NoStop}%
\bibitem [{Note3()}]{Note3}%
  \BibitemOpen
  \bibinfo {note} {NRST: https://roman.gsfc.nasa.gov}\BibitemShut {NoStop}%
\bibitem [{\citenamefont {DeRose}\ \emph {et~al.}(2021)\citenamefont {DeRose}
  \emph {et~al.}}]{DES_Y3_3x2pt_MC}%
  \BibitemOpen
  \bibfield  {author} {\bibinfo {author} {\bibfnamefont {J.}~\bibnamefont
  {DeRose}} \emph {et~al.} (\bibinfo {collaboration} {DES}),\ }\href@noop {}
  {\bibfield  {journal} {\bibinfo  {journal} {arXiv:2105.13547}\ } (\bibinfo
  {year} {2021})}\BibitemShut {NoStop}%
\bibitem [{\citenamefont {Hildebrandt}\ \emph {et~al.}(2017)\citenamefont
  {Hildebrandt} \emph {et~al.}}]{KiDS_modify}%
  \BibitemOpen
  \bibfield  {author} {\bibinfo {author} {\bibfnamefont {H.}~\bibnamefont
  {Hildebrandt}} \emph {et~al.} (\bibinfo {collaboration} {KiDS}),\ }\href@noop
  {} {\bibfield  {journal} {\bibinfo  {journal} {MNRAS}\ }\textbf {\bibinfo
  {volume} {465}},\ \bibinfo {pages} {1454} (\bibinfo {year}
  {2017})}\BibitemShut {NoStop}%
\bibitem [{\citenamefont {Joachimi}\ \emph {et~al.}(2021)\citenamefont
  {Joachimi} \emph {et~al.}}]{KiDS-1000_2021_MC}%
  \BibitemOpen
  \bibfield  {author} {\bibinfo {author} {\bibfnamefont {B.}~\bibnamefont
  {Joachimi}} \emph {et~al.} (\bibinfo {collaboration} {KiDS}),\ }\href@noop {}
  {\bibfield  {journal} {\bibinfo  {journal} {A\&A}\ }\textbf {\bibinfo
  {volume} {646}},\ \bibinfo {pages} {A129} (\bibinfo {year}
  {2021})}\BibitemShut {NoStop}%
\bibitem [{\citenamefont {Krause}\ \emph {et~al.}(2017)\citenamefont {Krause}
  \emph {et~al.}}]{Krause_2017}%
  \BibitemOpen
  \bibfield  {author} {\bibinfo {author} {\bibfnamefont {E.}~\bibnamefont
  {Krause}} \emph {et~al.},\ }\href@noop {} {\bibfield  {journal} {\bibinfo
  {journal} {arXiv:1706.09359v1}\ } (\bibinfo {year} {2017})}\BibitemShut
  {NoStop}%
\bibitem [{Note4()}]{Note4}%
  \BibitemOpen
  \bibinfo {note} {The studies presented in this paper required about six
  million hours of CPU time, 83\% of which was used in the ensemble
  tests}\BibitemShut {NoStop}%
\bibitem [{\citenamefont {Krause}\ \emph {et~al.}(2021)\citenamefont {Krause}
  \emph {et~al.}}]{DES_Y3_3x2pt_validation}%
  \BibitemOpen
  \bibfield  {author} {\bibinfo {author} {\bibfnamefont {E.}~\bibnamefont
  {Krause}} \emph {et~al.} (\bibinfo {collaboration} {DES}),\ }\href@noop {}
  {\bibfield  {journal} {\bibinfo  {journal} {arXiv:2105.13548}\ } (\bibinfo
  {year} {2021})}\BibitemShut {NoStop}%
\bibitem [{\citenamefont {Abbott}\ \emph {et~al.}(2018)\citenamefont {Abbott}
  \emph {et~al.}}]{DES_Y1_PRD}%
  \BibitemOpen
  \bibfield  {author} {\bibinfo {author} {\bibfnamefont {T.~M.~C.}\
  \bibnamefont {Abbott}} \emph {et~al.} (\bibinfo {collaboration} {DES}),\
  }\href@noop {} {\bibfield  {journal} {\bibinfo  {journal} {Phys.\ Rev. D}\
  }\textbf {\bibinfo {volume} {98}},\ \bibinfo {pages} {043526} (\bibinfo
  {year} {2018})}\BibitemShut {NoStop}%
\bibitem [{\citenamefont {Zuntz}\ \emph {et~al.}(2015)\citenamefont {Zuntz},
  \citenamefont {Paterno}, \citenamefont {Jennings}, \citenamefont {Rudd},
  \citenamefont {Manzotti}, \citenamefont {Dodelson}, \citenamefont {Bridle},
  \citenamefont {Sehrish},\ and\ \citenamefont {Kowalkowski}}]{Cosmosis_paper}%
  \BibitemOpen
  \bibfield  {author} {\bibinfo {author} {\bibfnamefont {J.}~\bibnamefont
  {Zuntz}}, \bibinfo {author} {\bibfnamefont {M.}~\bibnamefont {Paterno}},
  \bibinfo {author} {\bibfnamefont {E.}~\bibnamefont {Jennings}}, \bibinfo
  {author} {\bibfnamefont {D.}~\bibnamefont {Rudd}}, \bibinfo {author}
  {\bibfnamefont {A.}~\bibnamefont {Manzotti}}, \bibinfo {author}
  {\bibfnamefont {S.}~\bibnamefont {Dodelson}}, \bibinfo {author}
  {\bibfnamefont {S.}~\bibnamefont {Bridle}}, \bibinfo {author} {\bibfnamefont
  {S.}~\bibnamefont {Sehrish}},\ and\ \bibinfo {author} {\bibfnamefont
  {J.}~\bibnamefont {Kowalkowski}},\ }\href@noop {} {\bibfield  {journal}
  {\bibinfo  {journal} {Astronomy and Computing}\ }\textbf {\bibinfo {volume}
  {12}},\ \bibinfo {pages} {45} (\bibinfo {year} {2015})}\BibitemShut {NoStop}%
\bibitem [{Cos({\natexlab{a}})}]{Cosmosis_program}%
  \BibitemOpen
  \href@noop {} {\bibinfo {title} {To run cosmosis use the installer available
  at https://bitbucket.org/joezuntz/cosmosis/wiki/home.}}
  ({\natexlab{a}})\BibitemShut {NoStop}%
\bibitem [{\citenamefont {Hirata}\ and\ \citenamefont
  {Seljak}(2004)}]{Hirata-Seljak}%
  \BibitemOpen
  \bibfield  {author} {\bibinfo {author} {\bibfnamefont {C.~M.}\ \bibnamefont
  {Hirata}}\ and\ \bibinfo {author} {\bibfnamefont {U.}~\bibnamefont
  {Seljak}},\ }\href@noop {} {\bibfield  {journal} {\bibinfo  {journal} {Phys.
  Rev. D}\ }\textbf {\bibinfo {volume} {70}},\ \bibinfo {pages} {063526}
  (\bibinfo {year} {2004})}\BibitemShut {NoStop}%
\bibitem [{\citenamefont {Hirata}\ and\ \citenamefont
  {Seljak}(2010)}]{Hirata-Seljak-erratum}%
  \BibitemOpen
  \bibfield  {author} {\bibinfo {author} {\bibfnamefont {C.~M.}\ \bibnamefont
  {Hirata}}\ and\ \bibinfo {author} {\bibfnamefont {U.}~\bibnamefont
  {Seljak}},\ }\href@noop {} {\bibfield  {journal} {\bibinfo  {journal} {Phys.
  Rev. D}\ }\textbf {\bibinfo {volume} {82}},\ \bibinfo {pages} {049901(E)}
  (\bibinfo {year} {2010})}\BibitemShut {NoStop}%
\bibitem [{\citenamefont {Bridle}\ and\ \citenamefont
  {King}(2007)}]{Bridle-King_IA}%
  \BibitemOpen
  \bibfield  {author} {\bibinfo {author} {\bibfnamefont {S.}~\bibnamefont
  {Bridle}}\ and\ \bibinfo {author} {\bibfnamefont {L.}~\bibnamefont {King}},\
  }\href@noop {} {\bibfield  {journal} {\bibinfo  {journal} {New Journal of
  Physics}\ }\textbf {\bibinfo {volume} {9}},\ \bibinfo {pages} {444} (\bibinfo
  {year} {2007})}\BibitemShut {NoStop}%
\bibitem [{\citenamefont {Troxel}\ \emph {et~al.}(2018)\citenamefont {Troxel}
  \emph {et~al.}}]{Troxel_2018}%
  \BibitemOpen
  \bibfield  {author} {\bibinfo {author} {\bibfnamefont {M.~A.}\ \bibnamefont
  {Troxel}} \emph {et~al.},\ }\href@noop {} {\bibfield  {journal} {\bibinfo
  {journal} {Phys. Rev. D}\ }\textbf {\bibinfo {volume} {98}},\ \bibinfo
  {pages} {043528} (\bibinfo {year} {2018})}\BibitemShut {NoStop}%
\bibitem [{\citenamefont {Lewis}\ \emph {et~al.}(2000)\citenamefont {Lewis},
  \citenamefont {Challinor},\ and\ \citenamefont {Lasenby}}]{CAMB}%
  \BibitemOpen
  \bibfield  {author} {\bibinfo {author} {\bibfnamefont {A.}~\bibnamefont
  {Lewis}}, \bibinfo {author} {\bibfnamefont {A.}~\bibnamefont {Challinor}},\
  and\ \bibinfo {author} {\bibfnamefont {A.}~\bibnamefont {Lasenby}},\
  }\href@noop {} {\bibfield  {journal} {\bibinfo  {journal} {The Astrophysical
  Journal}\ }\textbf {\bibinfo {volume} {538}},\ \bibinfo {pages} {473}
  (\bibinfo {year} {2000})}\BibitemShut {NoStop}%
\bibitem [{\citenamefont {Smith}\ \emph {et~al.}(2003)\citenamefont {Smith},
  \citenamefont {Peacock}, \citenamefont {A.~Jenkins}, \citenamefont {Frenk},
  \citenamefont {Pearce}, \citenamefont {Thomas}, \citenamefont {Efstathiou},\
  and\ \citenamefont {Couchman}}]{Halofit_1}%
  \BibitemOpen
  \bibfield  {author} {\bibinfo {author} {\bibfnamefont {R.~E.}\ \bibnamefont
  {Smith}}, \bibinfo {author} {\bibfnamefont {J.~A.}\ \bibnamefont {Peacock}},
  \bibinfo {author} {\bibfnamefont {S.~D. M.~W.}\ \bibnamefont {A.~Jenkins}},
  \bibinfo {author} {\bibfnamefont {C.~S.}\ \bibnamefont {Frenk}}, \bibinfo
  {author} {\bibfnamefont {F.~R.}\ \bibnamefont {Pearce}}, \bibinfo {author}
  {\bibfnamefont {P.~A.}\ \bibnamefont {Thomas}}, \bibinfo {author}
  {\bibfnamefont {G.}~\bibnamefont {Efstathiou}},\ and\ \bibinfo {author}
  {\bibfnamefont {H.~M.~P.}\ \bibnamefont {Couchman}},\ }\href@noop {}
  {\bibfield  {journal} {\bibinfo  {journal} {Mon. Not. R. Astron. Soc.}\
  }\textbf {\bibinfo {volume} {341}},\ \bibinfo {pages} {1311} (\bibinfo {year}
  {2003})}\BibitemShut {NoStop}%
\bibitem [{\citenamefont {Bird}\ \emph {et~al.}(2012)\citenamefont {Bird},
  \citenamefont {Viel},\ and\ \citenamefont {Haehnelt}}]{Halofit_2}%
  \BibitemOpen
  \bibfield  {author} {\bibinfo {author} {\bibfnamefont {S.}~\bibnamefont
  {Bird}}, \bibinfo {author} {\bibfnamefont {M.}~\bibnamefont {Viel}},\ and\
  \bibinfo {author} {\bibfnamefont {M.~G.}\ \bibnamefont {Haehnelt}},\
  }\href@noop {} {\bibfield  {journal} {\bibinfo  {journal} {Mon. Not. R.
  Astron. Soc.}\ }\textbf {\bibinfo {volume} {420}},\ \bibinfo {pages} {2551}
  (\bibinfo {year} {2012})}\BibitemShut {NoStop}%
\bibitem [{\citenamefont {Takahashi}\ \emph {et~al.}(2012)\citenamefont
  {Takahashi}, \citenamefont {Sato}, \citenamefont {Nishimichi}, \citenamefont
  {Taruya},\ and\ \citenamefont {Oguri}}]{Halofit_3}%
  \BibitemOpen
  \bibfield  {author} {\bibinfo {author} {\bibfnamefont {R.}~\bibnamefont
  {Takahashi}}, \bibinfo {author} {\bibfnamefont {M.}~\bibnamefont {Sato}},
  \bibinfo {author} {\bibfnamefont {T.}~\bibnamefont {Nishimichi}}, \bibinfo
  {author} {\bibfnamefont {A.}~\bibnamefont {Taruya}},\ and\ \bibinfo {author}
  {\bibfnamefont {M.}~\bibnamefont {Oguri}},\ }\href@noop {} {\bibfield
  {journal} {\bibinfo  {journal} {The Astrophysical Journal}\ }\textbf
  {\bibinfo {volume} {761}},\ \bibinfo {pages} {152} (\bibinfo {year}
  {2012})}\BibitemShut {NoStop}%
\bibitem [{Cos({\natexlab{b}})}]{CosmoSIS_branch}%
  \BibitemOpen
  \href@noop {} {\bibinfo {title} {Used cosmosis version v1.6, dated
  2019-07-09, which can be found as branch v1.6 in
  \url{https://bitbucket.org/joezuntz/cosmosis/commits/bed016fb4ea69187a072db1fb4f2ea3d57a57f76}.}}
  ({\natexlab{b}})\BibitemShut {NoStop}%
\bibitem [{\citenamefont {Feroz}\ \emph {et~al.}(2009)\citenamefont {Feroz},
  \citenamefont {Hobson},\ and\ \citenamefont {Bridges}}]{Multinest_ref}%
  \BibitemOpen
  \bibfield  {author} {\bibinfo {author} {\bibfnamefont {F.}~\bibnamefont
  {Feroz}}, \bibinfo {author} {\bibfnamefont {M.~P.}\ \bibnamefont {Hobson}},\
  and\ \bibinfo {author} {\bibfnamefont {M.}~\bibnamefont {Bridges}},\
  }\href@noop {} {\bibfield  {journal} {\bibinfo  {journal} {Mon. Not. R.
  Astron. Soc.}\ }\textbf {\bibinfo {volume} {398}},\ \bibinfo {pages} {1601}
  (\bibinfo {year} {2009})}\BibitemShut {NoStop}%
\bibitem [{Note5()}]{Note5}%
  \BibitemOpen
  \bibinfo {note} {Increasing the value of $tolerance$ from 0.1 to $10^{-3}$
  ($10^{-9}$) increases the CPU time by a factor of 1.8 (10.8).}\BibitemShut
  {Stop}%
\bibitem [{\citenamefont {{Hinton}}(2016)}]{Chainconsumer_ref}%
  \BibitemOpen
  \bibfield  {author} {\bibinfo {author} {\bibfnamefont {S.~R.}\ \bibnamefont
  {{Hinton}}},\ }\bibfield  {title} {\bibinfo {title} {{ChainConsumer}},\
  }\href@noop {} {\bibfield  {journal} {\bibinfo  {journal} {The Journal of
  Open Source Software}\ }\textbf {\bibinfo {volume} {1}},\ \bibinfo {eid}
  {00045} (\bibinfo {year} {2016})},\ \bibinfo {note} {(See also
  \url{https://samreay.github.io/ChainConsumer/})}\BibitemShut {NoStop}%
\bibitem [{KDE()}]{KDE_algo}%
  \BibitemOpen
  \href@noop {} {\bibinfo {title} {See for example
  \url{https://en.wikipedia.org/wiki/Kernel_density_estimation}.}}\BibitemShut
  {Stop}%
\bibitem [{\citenamefont {Zyla}\ \emph {et~al.}(2020)\citenamefont {Zyla} \emph
  {et~al.}}]{Zyla:2020zbs}%
  \BibitemOpen
  \bibfield  {author} {\bibinfo {author} {\bibfnamefont {P.}~\bibnamefont
  {Zyla}} \emph {et~al.} (\bibinfo {collaboration} {Particle Data Group}),\
  }\bibfield  {title} {\bibinfo {title} {{Review of Particle Physics}},\ }\href
  {https://doi.org/10.1093/ptep/ptaa104} {\bibfield  {journal} {\bibinfo
  {journal} {PTEP}\ }\textbf {\bibinfo {volume} {2020}},\ \bibinfo {pages}
  {083C01} (\bibinfo {year} {2020})},\ \bibinfo {note} {(See also
  \url{https://pdg.lbl.gov})}\BibitemShut {NoStop}%
\bibitem [{\citenamefont {Eadie}\ \emph {et~al.}(1971)\citenamefont {Eadie},
  \citenamefont {Drijard}, \citenamefont {James}, \citenamefont {Roos},\ and\
  \citenamefont {Sadoulet}}]{pull_reference}%
  \BibitemOpen
  \bibfield  {author} {\bibinfo {author} {\bibfnamefont {W.~T.}\ \bibnamefont
  {Eadie}}, \bibinfo {author} {\bibfnamefont {D.}~\bibnamefont {Drijard}},
  \bibinfo {author} {\bibfnamefont {F.}~\bibnamefont {James}}, \bibinfo
  {author} {\bibfnamefont {M.}~\bibnamefont {Roos}},\ and\ \bibinfo {author}
  {\bibfnamefont {B.}~\bibnamefont {Sadoulet}},\ }\href@noop {} {\emph
  {\bibinfo {title} {Statistical Methods in Experimental Physics}}}\ (\bibinfo
  {publisher} {North Holland},\ \bibinfo {year} {1971})\ pp.\ \bibinfo {pages}
  {277--278}\BibitemShut {NoStop}%
\bibitem [{\citenamefont {Demortier}\ and\ \citenamefont
  {Lyons}(2002)}]{pull_LucDemortier}%
  \BibitemOpen
  \bibfield  {author} {\bibinfo {author} {\bibfnamefont {L.}~\bibnamefont
  {Demortier}}\ and\ \bibinfo {author} {\bibfnamefont {L.}~\bibnamefont
  {Lyons}},\ }\href@noop {} {\emph {\bibinfo {title} {Everything you always
  wanted to know about pulls}}}\ (\bibinfo {year} {2002})\ \bibinfo {note}
  {(CDF/ANAL/PUBLIC/-5776)}\BibitemShut {NoStop}%
\bibitem [{Stu()}]{Studies_1-2x2pt}%
  \BibitemOpen
  \href@noop {} {\bibinfo {title} {These studies are currently under
  way.}}\BibitemShut {Stop}%
\end{thebibliography}%

\end{document}